\documentclass{nature2}

\usepackage{graphicx}        
\usepackage[utf8]{inputenc}
\usepackage{hyperref}
\usepackage{wasysym}
\usepackage{bookmark}
\usepackage{longtable}
\usepackage{bm}
\usepackage[british]{babel}
\usepackage{color}

\usepackage{bibunits}
\usepackage{etoolbox}
\usepackage{lipsum}
\usepackage{lscape}

\defaultbibliographystyle{naturemag}

\definecolor{orange}{RGB}{255,127,0}
\definecolor{darkgreen}{RGB}{8,114,11}

\hypersetup{
	hidelinks = true,
}

\makeatletter
\newcommand{\ion}{}
\DeclareRobustCommand*{\ion}[2]{%
  #1$\;$%
  \if b\expandafter\@car\f@series\relax\@nil
    \begingroup 
      \sbox0{\rmfamily\mdseries\textsc{v}}%
      \resizebox{!}{\ht0}{\rmfamily\@Roman{#2}}%
    \endgroup
  \else
    \textsc{\rmfamily\@roman{#2}}%
  \fi
}
\makeatother

\makeatletter
\newcounter{mybibcounter}

\makeatother

\def\kms{km s$^{-1}$}         
\def\ms{\hbox{m s$^{-1}$}}         
\def\gcm3{\hbox{g cm$^{-3}$}}       
\def\vsini{\hbox{$\upsilon \sin i_{\star}$}}      
\def\Msun{\hbox{M$_{\astrosun}$}}             
\def\Rsun{\hbox{R$_{\astrosun}$}}

\def\Mearth{\hbox{M$_{\oplus}$}}
\def\Rearth{\hbox{R$_{\oplus}$}}
\def\degr{\hbox{$^\circ$}}
\def\teff{T$_{\rm eff}$}
\def\logg{log~{\it g}}
\def\met{[Fe/H]}

\title{An extremely low-density and temperate giant exoplanet}

\author{A.~Santerne$^{\ref{LAM}}$, 
	L.~Malavolta$^{\ref{catania}}$, 
	M.~R.~Kosiarek$^{\ref{UCSC},\ref{NSF}}$, 
	F.~Dai$^{\ref{Kalvi}, \ref{princeton}, \ref{DGPS}}$, 
	C.~D.~Dressing$^{\ref{UCB}}$, 
	X.~Dumusque$^{\ref{geneve}}$, 
	N.~C.~Hara$^{\ref{geneve}, \ref{CHEOPS}}$, 
	T.~A.~Lopez$^{\ref{LAM}}$,
	A.~Mortier$^{\ref{cambridge}}$, 
	A.~Vanderburg$^{\ref{austin}}$, 
	V. ~Adibekyan$^{\ref{CAUP}}$, 
	D.~J.~Armstrong$^{\ref{warwick1}, \ref{warwick2}}$, 
	D.~Barrado$^{\ref{CAB}}$, 
	S.~C.~C.~Barros$^{\ref{CAUP}}$, 
	D.~Bayliss$^{\ref{warwick2}}$, 
	D.~Berardo$^{\ref{MIT}}$, 
	I.~Boisse$^{\ref{LAM}}$, 
	A.~S.~Bonomo$^{\ref{torino}}$, 
	F.~Bouchy$^{\ref{geneve}}$, 
	D.~J.~A.~Brown$^{\ref{warwick1}, \ref{warwick2}}$, 
	L.~A.~Buchhave$^{\ref{DTU}}$, 
	R.~P.~Butler$^{\ref{washington}}$, 
	A.~Collier~Cameron$^{\ref{StAndrews}}$, 
	R.~Cosentino$^{\ref{TNG}}$, 
	J.~D.~Crane$^{\ref{carnegie}}$, 
	I.~J.~M.~Crossfield$^{\ref{Kalvi}, \ref{UKansas}}$, 
	M.~Damasso$^{\ref{torino}}$, 
	M.~R.~Deleuil$^{\ref{LAM}}$, 
	E.~Delgado~Mena$^{\ref{CAUP}}$, 
	O.~Demangeon$^{\ref{CAUP}}$, 
	R.~F.~D\'iaz$^{\ref{UBA}, \ref{CONICET}}$, 
	J.-F.~Donati$^{\ref{IRAP}}$, 
	P.~Figueira$^{\ref{eso}, \ref{CAUP}}$, 
	B.~J.~Fulton$^{\ref{IPAC}}$,
	A.~Ghedina$^{\ref{TNG}}$, 
	A.~Harutyunyan$^{\ref{TNG}}$, 
	G.~H\'ebrard$^{\ref{IAP}}$, 
	L.~A.~Hirsch$^{\ref{standford}}$, 
	S.~Hojjatpanah$^{\ref{CAUP},\ref{UPorto}}$, 
	A.~W.~Howard$^{\ref{caltech}}$, 
	H.~Isaacson$^{\ref{UCB}}$, 
	D.~W.~Latham$^{\ref{CFA}}$, 
	J.~Lillo-Box$^{\ref{CAB}, \ref{eso}}$, 
	M.~L\'opez-Morales$^{\ref{CFA}}$, 
	C.~Lovis$^{\ref{geneve}}$, 
	A.~F.~Martinez~Fiorenzano$^{\ref{TNG}}$, 
	E.~Molinari$^{\ref{Cagliari}}$, 
	O.~Mousis$^{\ref{LAM}}$, 
	C.~Moutou$^{\ref{IRAP}}$, 
	C.~Nava$^{\ref{CFA}}$, 
	L.~D.~Nielsen$^{\ref{geneve}}$, 
	H.~P.~Osborn$^{\ref{LAM}}$, 
	E.~A.~Petigura$^{\ref{UCLA}}$, 
	D.~F.~Phillips$^{\ref{CFA}}$, 
	D.~L.~Pollacco$^{\ref{warwick1}, \ref{warwick2}}$, 
	E.~Poretti$^{\ref{TNG}}$, 
	K.~Rice$^{\ref{SUPA}, \ref{Edinburg}}$, 
	N.~C.~Santos.$^{\ref{CAUP},\ref{UPorto}}$, 
	D.~S\'egransan$^{\ref{geneve}}$, 
	S.~A.~Shectman$^{\ref{carnegie}}$, 
	E.~Sinukoff$^{\ref{caltech}}$, 
	S.~G.~Sousa$^{\ref{CAUP}}$, 
	A.~Sozzetti$^{\ref{torino}}$, 
	J.~K.~Teske$^{\ref{carnegie}, \ref{Hubble}}$, 
	S.~Udry$^{\ref{geneve}}$, 
	A.~Vigan$^{\ref{LAM}}$, 
	S.~X.~Wang$^{\ref{carnegie}}$, 
	C.~A.~Watson$^{\ref{Belfast}}$, 
	L.~M.~Weiss$^{\ref{hawaii}}$, 
	P.~J.~Wheatley$^{\ref{warwick1}, \ref{warwick2}}$, 
	J.~N.~Winn$^{\ref{princeton}}$}

\begin{document}

\maketitle

\begin{affiliations}
\item Aix Marseille Univ, CNRS, CNES, LAM, Marseille, France\label{LAM}
\item INAF - Osservatorio Astrofisico di Catania, Via S.Sofia 78, 95123 Catania, Italy\label{catania}
\item Department of Astronomy and Astrophysics, University of California, Santa Cruz, CA 95064, USA\label{UCSC}
\item NSF Graduate Research Fellow\label{NSF}
\item Department of Physics and Kavli Institute for Astrophysics and Space Research, Massachusetts Institute of Technology, Cambridge, MA, 02139, USA\label{Kalvi}
\item Department of Astrophysical Sciences, Princeton University, 4 Ivy Lane, Princeton, NJ 08544 USA\label{princeton}
\item Division of Geological and Planetary Sciences, California Institute of Technology,1200 East California Blvd, Pasadena, CA, USA 91125\label{DGPS}
\item Department of Astronomy, University of California, Berkeley, Berkeley, CA 94720, USA\label{UCB}
\item Department of Astronomy of the University of Geneva, 51 Chemin des Maillettes, 1290 Versoix, Switzerland\label{geneve}
\item CHEOPS fellow\label{CHEOPS}
\item Cavendish Laboratory and Kavli Institute for Cosmology, University of Cambridge, J.J. Thomson Avenue, Cambridge CB3 0HE, UK\label{cambridge}
\item Department of Astronomy, The University of Texas at Austin, Austin, TX 78712, USA\label{austin}
\item Instituto de Astrof\'isica e Ci\^encias do Espa\c{c}co, Universidade do Porto, CAUP, Rua das Estrelas, 4150-762 Porto, Portugal\label{CAUP}
\item Centre for Exoplanets and Habitability, University of Warwick, Gibbet Hill Road, Coventry, CV4 7AL, UK\label{warwick1}
\item Department of Physics, University of Warwick, Gibbet Hill Road, Coventry, CV4 7AL, UK\label{warwick2}
\item Centro de Astrobiolog\'ia (CSIC-INTA), ESAC campus 28692 Villanueva de la Ca\~nada (Madrid), Spain\label{CAB}
\item Department of Physics, and Kavli Institute for Astrophysics and Space Research, Massachusetts Institute of Technology, Cambridge, MA, USA\label{MIT}
\item INAF - Osservatorio Astrofisico di Torino, via Osservatorio 20, 10025 Pino Torinese, Italy\label{torino}
\item DTU Space, National Space Institute, Technical University of Denmark, Elektrovej 328, DK-2800 Kgs. Lyngby, Denmark\label{DTU}
\item Department of Terrestrial Magnetism, Carnegie Institution for Science, 5241 Broad Branch Road NW, Washington DC 20015\label{washington}
\item Centre for Exoplanet Science, SUPA School of Physics and Astronomy, University of St Andrews, North Haugh, St Andrews KY16 9SS, UK\label{StAndrews}
\item INAF - Fundacion Galileo Galilei, Rambla Jos\`e Ana Fernandez P\`erez, 7 - Bre\~na Baja, TF - Spain\label{TNG}
\item The University of Kansas, Department of Physics and Astronomy, Malott Room 1082, 1251 Wescoe Hall Drive, Lawrence, KS, 66045, USA\label{UKansas}
\item The Observatories of the Carnegie Institution for Science, 813 Santa Barbara Street, Pasadena, CA, USA 91101\label{carnegie}
\item Universidad de Buenos Aires, Facultad de Ciencias Exactas y Naturales, Buenos Aires, Argentina\label{UBA}
\item CONICET - Universidad de Buenos Aires. Instituto de Astronom\'ia y F\'isica del Espacio (IAFE), Buenos Aires, Argentina\label{CONICET}
\item IRAP, 14 avenue Edouard Belin 31400 Toulouse, France\label{IRAP}
\item European Southern Observatory, Alonso de Cordova 3107, Vitacura, Santiago, Chile\label{eso}
\item NASA Exoplanet Science Institute / Caltech-IPAC, Pasadena, CA, USA 91106\label{IPAC}
\item Institut d'Astrophysique de Paris, UMR7095 CNRS, Universit\'e Pierre \& Marie Curie, 98bis boulevard Arago, 75014 Paris, France\label{IAP}
\item Kavli Institute for Particle Astrophysics and Cosmology, Stanford University, Stanford, CA, USA\label{standford}
\item Departamento de Fisica e Astronomia, Faculdade de Ci\^{e}ncias, Universidade do Porto, Rua Campo Alegre, 4169-007 Porto, Portugal\label{UPorto}
\item California Institute of Technology, Pasadena, CA 91125, USA\label{caltech}
\item Center for Astrophysics ${\rm \mid}$ Harvard {\rm \&} Smithsonian, 60 Garden Street, Cambridge, MA 02138, USA\label{CFA}
\item INAF Osservatorio Astronomico di Cagliari \& REM - Via della Scienza 5 - 09047 Selargius CA, Italy\label{Cagliari}
\item Department of Physics \& Astronomy, University of California Los Angeles, Los Angeles, CA 90095, USA\label{UCLA}
\item SUPA, Institute for Astronomy, University of Edinburgh, Royal Observatory, Blackford Hill, Edinburgh, EH93HJ, UK\label{SUPA}
\item Centre for Exoplanet Science, University of Edinburgh, Edinburgh, UK\label{Edinburg}
\item  NASA Hubble Fellow\label{Hubble}
\item Astrophysics Research Centre, Queen's University Belfast, Belfast BT7 1NN, UK\label{Belfast}
\item Institute for Astronomy, 2680 Woodlawn Dr., Honolulu, HI 96822, USA\label{hawaii}
\end{affiliations}

\begin{bibunit}

\begin{abstract}
Transiting extrasolar planets are key objects in the study of the formation, migration, and evolution of planetary systems\cite{fortney2013framework}. In particular, the exploration of the atmospheres of giant planets, through transmission spectroscopy or direct imaging, has revealed a large diversity in their chemical composition and physical properties\cite{sing2016continuum}. Studying these giant planets allows one to test the global climate models\cite{charnay20153d,parmentier2016transitions} that are used for the Earth and other solar system planets. However, these studies are mostly limited either to highly-irradiated transiting giant planets or directly-imaged giant planets at large separations. Here we report the physical characterisation of the planets in a bright multi-planetary system (HIP41378) in which the outer planet, HIP41378 $f$ is a Saturn-sized planet (9.2$\pm$0.1 \Rearth) with an anomalously low density of 0.09$\pm$0.02 \gcm3\ that is not yet understood. Its equilibrium temperature is about 300 K. Therefore, it represents a planet with a mild temperature, in between the hot Jupiters and the colder giant planets of the Solar System. It opens a new window for atmospheric characterisation of giant exoplanets with a moderate irradiation, with the next-generation space telescopes such as $JWST$\cite{greene2016characterizing} and $ARIEL$\cite{tinetti2016science} as well as the extremely-large ground-based telescopes. HIP41378 $f$ is thus an important laboratory to understand the effect of the irradiation on the physical properties and chemical composition of the atmosphere of planets.
\end{abstract}

The star HIP41378 was initially observed by the $Kepler$ space telescope\cite{borucki2010kepler} during campaign C5 of the $K2$ mission\cite{howell2014k2} from 2015-04-27 to 2015-07-10. The data revealed 5 transiting planets\cite{vanderburg2016five}. Two of them have orbital periods of about 15 and 31 days and are named planets $b$ and $c$, respectively. They transited several times within the 80 continuous days of space-based photometry. The other 3 planets, named $d$, $e$, and $f$, exhibit only one transit during C5. From their transit durations, these planets were predicted to have orbital periods of several months to years. 

HIP41378 was re-observed during K2 campaign C18 from 2018-05-12 to 2018-07-02 enabling the detection of new transits of the innermost planets as well as a second transit of planets $d$ and $f$. The new data permitted the derivation of a set of possible ephemerides for the outer planets\cite{becker2018discrete,berardo2019revisiting} with periods of about 1000 days and all harmonics down to about 50 days. The $TESS$ space telescope\cite{ricker2010transiting} also observed this system from 2019-01-07 to 2019-02-02 and detected one transit of each of the two innermost planets. Planets $b$ to $e$ have radii smaller than 5 Earth radii (\Rearth) and were expected to be of relatively low mass. However, planet $f$ is a Saturn-sized planet and was therefore expected to exhibit a relatively large radial-velocity signal$^{\citenum{vanderburg2016five}}$, at the level of about 30 \ms. 

To measure the planets' masses and refine their orbital periods and properties, the host star was subsequently observed with precise radial-velocity spectrographs. The first observations with the SOPHIE spectrograph on the 1.93-m telescope at Haute-Provence Observatory (France) revealed no significant variation, indicating that planet $f$ is likely of lower mass than predicted. The star was then observed by various high-precision radial-velocity facilities: the HARPS, HARPS-N, HIRES, and the PFS spectrographs (see Methods). A total of 464 radial-velocity nightly-binned measurements have been collected over four years with these 4 instruments (see Methods and Supplementary Tables 2 -- 5). They are displayed in Figure 1.

The host star HIP41378 is a late-F type dwarf with an effective temperature of \teff = 6290$\pm$77 K and an iron abundance of \met = -0.05 $\pm$ 0.10 dex (Lund et al., submitted). The star exhibits a low activity signal with a photometric amplitude of 200 ppm and a period of 6.4$\pm$0.8 days (although it might be twice this value, see Methods). This activity signal is not observed in spectroscopy (See Methods and Supplementary Figure 2). The star was also observed in spectropolarimetry with the ESPaDOnS spectropolarimeter and no magnetic field was detected confirming its low-activity level (see Methods). Thanks to high-precision and short-cadence photometric observations during C18, asteroseismic constraints on the fundamental properties of the host star were derived (Lund et al., submitted). In particular, the mean stellar density of the star was measured to be 0.785$\pm$0.008 \gcm3.

In order to refine the orbital periods of planets $d$, $e$, and $f$, the precise radial velocities from the four aforementioned instruments were first analysed without the photometric data (see Methods). The results of this preliminary analysis revealed two main aspects. Firstly, the orbital period of planet $f$ is only compatible with the 542-d solution allowed by the $K2$ photometry (See Supplementary Figure 4). This solution is also the one that minimises the orbital eccentricity of this planet$^{\citenum{berardo2019revisiting}}$. Secondly, there is a significant signal for a non-transiting planet with a period of about 62 days (see Supplementary Figure 4). This new planet, called $g$, is in a 2:1 mean-motion resonance (MMR) with planet $c$ and might explain its observed $\sim$ 85-minutes transit timing variations$^{\citenum{berardo2019revisiting}}$. It is not possible for this 62-d signal to be either the planets $d$ or $e$ as they would need a high eccentricity (greater than $\sim$0.6) to explain their long transit duration$^{\citenum{berardo2019revisiting}}$, and the system would likely be unstable$^{\citenum{becker2018discrete}}$. In this preliminary analysis, planets $d$ and $e$ are not significantly detected.

The radial-velocity data were then jointly analysed with the photometric data from the two $K2$ campaigns with \texttt{PASTIS} (Planet Analysis and Small Transit Investigation Software)\cite{diaz2014pastis}. The SOPHIE data were not included in the analysis as they are not precise enough. The $TESS$ data were also excluded from the analysis because only one transit of planets $b$ and $c$ were detected during the sector 7 and none for the outer planets. The central star was modelled using stellar evolution tracks and atmosphere models (see Methods) taking into account the asteroseismic constraints provided by the $K2$ C18 photometry. Given that planet $d$ is undetected in the radial-velocity data, the 278-d solution for its orbital period was assumed in the analysis as it corresponds to the solution minimising its eccentricity (Lund et al., submitted).

The results are reported in Supplementary Table 7 and the physical parameters of the system are listed in Table \ref{Table1}. The radial-velocity signals of planets $b$, $c$, $g$, and $f$ are significantly detected allowing us to determine their planetary masses, 6.89$\pm$0.88 Earth masses (\Mearth), 4.4$\pm$1.1 \Mearth, 7.0$\pm$1.5 \Mearth, and 12$\pm$3 \Mearth, respectively. Since planet $g$ does not transit, its orbital inclination is unknown and thus its derived mass is actually a minimum mass. However, since the rest of the system is nearly coplanar, its inclination is expected to be close to $\sim$88\degr. Therefore, the actual mass of planet $g$ is expected to be nearly identical to the minimum mass derived here. The planets $d$ and $e$ are not significantly detected in the radial-velocity data indicating, with a 95\% credible probability, that their masses are less than 4.6 \Mearth\ and 22 \Mearth, respectively. Although only one transit of planet $e$ has been detected, its orbital period is constrained to 369$\pm$10 days thanks to the asteroseismic constraints (see Methods). As a consequence, planets $d$, $e$, and $f$ are near a 3:4:6 MMR chain, hence the entire system may be in a 1:2:4:18:24:36 MMR chain.
One might therefore speculate that more planets are still to be discovered in between HIP41378 $g$ and $d$ (see Supplementary Figure 6).

All the five transiting planets in this system have relatively low densities (see Table 1) and are therefore gaseous planets\cite{brugger2017constraints} (see Figure 2). The most extreme planet in this system is planet $f$ that exhibits a mass of 12$\pm$3 \Mearth\ for a radius of 9.2$\pm$0.1 \Rearth, leading to a bulk density of 0.09$\pm$0.02 \gcm3. A comparison of the physical properties of HIP41378 $f$ with theoretical models of the structure of low-mass exoplanets\cite{baraffe2008structure} for a 3.1-Gyr old system revealed that this planet would need a sub-solar metallicity to explain its large radius given its low mass (see Figure 2 and Supplementary Figure 8). As a consequence, this planet is likely composed of a large atmosphere dominated by hydrogen and helium and a very small core. Such a low-density planet with an age of 3.1 Gyr is not predicted by the current formation and evolution models of exoplanets\cite{mordasini2012characterization} and it will be challenge for such models to explain its history.

One possible explanation for the large radius of HIP41378 $f$ given its relatively low mass is that the planet is surrounded by an optically-thick ring\cite{akinsanmi2018detecting}. The presence of rings would artificially enlarge the apparent radius of the planet, hence decreasing its apparent density. This hypothesis will be presented in a forthcoming paper (Akinsanmi et al., in prep.). Observing a transit of a ringed planet $f$ in the infrared, where the rings are expected to be optically thinner than in the $Kepler$ bandpass, could reveal a significantly smaller planet. Another explanation is that HIP41378 $f$ is a ``super-puff'' planet with an extended, outflowing atmosphere\cite{wang2019dusty}.

Planet HIP41378 $f$ orbits with a period of $\sim$ 542 days (1.5 years), hence is at a semi-major axis of 1.4 astronomical unit (au) with an eccentricity smaller than 0.035 (at 95\% credible probability -- see Supplementary Table 7). Its equilibrium temperature, assuming a zero Bond albedo, is 294$^{_{+3}}_{^{-1}}$\ K. This planet is thus at the inner edge of the conservative habitable zone\cite{kopparapu2013habitable}. Although this large, gaseous planet is not likely to be habitable, it might host habitable exo-moons. Given the brightness of the host star and the favorable orbital period and eccentricity of HIP41738 $f$, the planet is therefore one of the best planets to search for habitable exo-moons. Constraining the presence of exo-moons in this system could give important insight to test the formation theories of the Galilean moons\cite{ronnet2018saturn}.

It is important to stress here that the six planets in this system have radial-velocity amplitudes at the level of 1\ms\ or even below (see Supplementary Table 7). Such low-amplitude signals are at the limit of the current instrumental stability. As a consequence, the masses and densities derived in this paper might be affected by unknown systematics. Even in this case, the data exclude a large mass for these planets, including for planet $f$, at odds with predictions based on its Saturn-like radius$^{\citenum{vanderburg2016five}}$. The 95\% credible upper limit on its mass is 18.6 \Mearth, giving a 95\% credible upper limit on the density of 0.13 \gcm3. More radial velocities with improved stability spectrographs like ESPRESSO\cite{pepe2010espresso} on the ESO / VLT are needed to increase the accuracy of these detections, secure the mass of planets $d$ and $e$, and detect possible additional planets, especially at orbital periods between those of planets $g$ and $d$.

With relatively low densities, hence low surface gravities, the planets in the HIP41378 system are excellent targets for atmospheric characterisation. It will be possible to probe the atmosphere of several planets within the same system with upcoming instruments on $JWST^{\citenum{greene2016characterizing}}$, $ARIEL^{\citenum{tinetti2016science}}$, and ground-based extremely large telescopes, allowing for the direct comparison of their chemical composition and physical properties (see Supplementary Figure 7). In this context, planet $f$ is also of special interest as it receives a modest stellar insolation flux of $S_{f}$ = 1.7 erg cm$^{-2}$ s$^{-1}$, or 1.3 $S_{\oplus}$. It is therefore filling the gap between the highly-irradiated hot Jupiters and the cold, solar-system giants Jupiter and Saturn (see Figure 3). The latter are used as calibrators for atmosphere models of giant planets. Being a temperate giant planet, its transmission spectrum might reveal different properties and chemical species than the ones observed in the much hotter planets.

The planet HIP41378 $f$ is joining the small population of extremely low-density planets, such as Kepler-51 $d$\cite{masuda2014very} or Kepler-79 $d$\cite{jontof2014kepler} that were characterised by TTVs. These planets are however orbiting at shorter periods and are transiting much fainter stars, which challenge their atmospheric characterisation\cite{libby2019featureless}.

Thanks to the asteroseismic constraints on the host star (Lund et al., submitted), it is possible to derive the radius of the planets as well as the system's age with an exquisite precision. Such a level of precision is the goal of ESA's $PLATO$ mission\cite{rauer2014plato}. Moreover, this is the first time that a low-mass exoplanet in the habitable zone of a solar-like star has been characterised, hence paving the way towards the characterisation of transiting true Earth analogs that will be detected by $PLATO$. HIP41378 is therefore a foretaste of the systems that this upcoming space mission will provide, in terms of both covered orbital period and precision on the fundamental parameters.

\putbib[HIP41378]

\end{bibunit}

\newpage

\begin{bibunit} 
\begin{methods}
\section{Observation and data reduction}

\subsection{K2}

The $K2$ long-cadence data from the C5 campaign and the short-cadence data from the C18 campaign were both reduced using the \texttt{K2SFF} pipeline\cite{vanderburg2014technique}. The self-flattening technique to correct for the systematics was applied taking into account the transit\cite{vanderburg2016planetary} to improve the light-curve precision preserving the transit shape and depth. The residual instrumental or stellar variability was then corrected using a spline with knots every 10 days.

\subsection{HARPS and HARPS-N}

The target star HIP41378 was monitored by the HARPS spectrograph\cite{phase2003setting} mounted on the 3.6-m telescope at the La Silla Observatory (Chile). A total of 370 spectra were collected over 3 years with a typical exposure time of 900s. These spectra have a signal-to-noise ratio (SNR) up to 140 per pixel at 550nm. The star was also observed by the HARPS-N spectrograph\cite{cosentino2012harps} mounted on the 3.6-m Italian Telescopio Nazionale Galileo (TNG) at the Roque de los Muchachos Observatory (La Palma, Spain). A total of 176 spectra were collected over 4 years with typical exposure times of 900s leading to SNR up to 170 per pixel at 550nm. The radial velocities for both instruments were reduced with their online pipeline which consists of cross-correlating\cite{baranne1996elodie, pepe2002coralie} the observed spectra with a binary mask corresponding to a G2 dwarf.

In order to test the high-frequency noise, such as granulation or p-modes for this late-F star, two intensive-cadence observations were performed with HARPS over two consecutive nights. Over each of these two nights (2018-03-10 and 2018-03-11) 24 spectra were collected continuously over 2.1 hours with exposure times of 300s (see Supplementary Figure 1). The two time series have an RMS of 3.7\ms\ for a photon noise of 3.7\ms. Since no excess of RMS is observed, the star is thus quiet over a timescale of hours and at the level of 1 -- 3 \ms\ (see Supplementary Figure 1). The HARPS observing strategy was initially to take 2 to 3 spectra each night separated by a few hours in order to average possible high-frequency noise\cite{dumusque2011planetary}. After this sequence, the strategy was changed to only one spectrum per night.

The number of spectra collected each night is highly heterogenous, between 1 and 24, and the instrumental calibrations are performed only once a night. Therefore, to avoid systematic effects with the daily calibrations, the HARPS and HARPS-N data were respectively nightly binned and a calibration noise of 0.5\ms\ was added quadratically to the radial-velocity uncertainties. Each night bin corresponds to an observation epoch. This leads to 216 HARPS and 155 HARPS-N epochs.

Given the low radial-velocity amplitudes of the signals, only the nights when the uncertainty is less than 5\ms\ were kept for the analysis. Finally, the end of life of the reference Thorium-Argon lamp on HARPS occurred on 2018-11-28. The data collected the two previous nights are affected by strong systematics and were rejected from the analysis.

The nightly-binned radial-velocity HARPS and HARPS-N data are available in the Supplementary Tables 2 and 3, respectively.

\subsection{HIRES}

A total of 218 radial velocities were obtained on HIP41378 with the High Resolution Echelle Spectrometer (HIRES)\cite{vogt1994hires} on the Keck I Telescope on Maunakea between 2016 September and 2019 May. These data were collected with the C2 decker with a typical SNR of 200 per pixel (250k on the exposure meter, $\sim$5-minute exposures). An iodine cell was used for wavelength calibration\cite{butler1996attaining}. Most of the HIRES data were collected with three consecutive exposures to better average over any stellar oscillations that might occur on short timescale. A higher resolution template observation was also collected with the B3 decker on 2016 October 10 with 1.1'' seeing. The template was a triple exposure with a total SNR of 340 per pixel (250k each on the exposure meter) without the iodine cell. The HIRES data collection, reduction, and analysis followed the methods of the California Planet Search\cite{howard2010occurrence}. Like for the other instruments, the 218 HIRES radial velocities were nightly binned to 75 epochs. They are listed in Supplementary Table 4.

\subsection{PFS}

Observations of HIP41378 were conducted with the Planet Finder Spectrograph (PFS)\cite{crane2006carnegie} on the 6.5m Magellan II telescope at Las Campanas Observatory in Chile in March-April 2016 and January 2017. PFS is an iodine-calibration precision RV spectrograph and all data are reduced and analysed by a custom IDL pipeline that has been shown to produce RVs with $<$ 1 m/s precision on bright, stable stars$^{\citenum{butler1996attaining}}$. The HIP41378 PFS observations were conducted on twenty different nights, often with multiple exposures per night to increase SNR per epoch. Twenty-seven individual iodine spectra were acquired in total with exposure times typically between 500 and 700 seconds and SNR values at peak blaze in the iodine orders typically between 120 and 190. These iodine spectra were taken in 1x1 binning mode with the 0.5x2.5'' slit, resulting in a resolving power $\sim$80,000; an iodine-free template spectrum (consisting of three consecutive exposures) was also obtained with the 0.3x.25'' slit, resulting in a higher resolving power $\sim$130,000. The final radial velocities are binned nightly and are reported in the Supplementary Table 5. 

\subsection{ESPaDOnS}

The star HIP41378 was also monitored with ESPaDOnS (Echelle SpectroPolarimetric Device for the Observation of Stars)\cite{donati2006espadons} over ten days. Eight polarimetric observations were obtained in late December 2017 and early January 2018, each of them consisting in a sequence of four consecutive spectra with exposure times between 600s and 840s, leading to SNR of 500 per pixel for the sequence. They were reduced with the same method as for the $\tau$ Bootis\cite{donati2008magnetic} which has similar stellar parameters to HIP41378 except for the rotation period (equal to 3.3d in the case of $\tau$ Bootis). No polarimetric signal is detected in the Stockes V data of HIP41378. The upper limit on the longitudinal magnetic field (i.e., the line-of-sight projected magnetic field averaged over the visible stellar hemisphere) is 1.5 G at 1-$\sigma$. This non-detection is consistent with the results on $\tau$ Bootis which is more active than HIP41378 and suggests that the surface magnetic field of HIP41378 is complex and does not exceed a few G (i.e. weaker than that of $\tau$ Bootis that reaches a maximum surface field of $\sim$10 G\cite{fares2009magnetic,mengel2016evolving}.

\section{Activity analysis in photometry and spectroscopy}

The photometric $K2$ data exhibit a faint rotational modulation with an amplitude at the level of 200 ppm. The autocorrelation function of the data indicates that the host star has a rotation period of 6.4$\pm$0.8 days (see Supplementary Figure 3). Combined with the stellar radius (see Table 1), this gives an equatorial velocity of $v_{\rm rot}$ = 10.1$\pm$1.3 \kms, significantly larger than the \vsini = 5.2$\pm$0.5 \kms\ measured on the HARPS spectra. This indicates that either the star has an inclination of $\sim$33$^\circ$ or that the true rotation period is twice the aforementionned one.
 
The periodogram of the HARPS and HARPS-N radial velocities, full width half maximum (FWHM), bisector, and \ion{Ca}{2} H\&K S index reveal no significant variability at the rotation period of the star. The activity index, derived from the S index\cite{noyes1984relation,lovis2011harps} is logR'$_{\rm H\&K}$ = -4.78$\pm$0.03. Therefore, the star HIP41378 is inactive at the precision of the spectroscopic data. Derived from the activity index, the rotation period of the star\cite{donahue1993surface,wright2004chromospheric} is 8$\pm$3 d and its age\cite{mamajek2008improved} is 2.52$\pm$0.23 Gyr. Both values are compatible with the ones derived from photometry and isochrones through the combined analysis, respectively, to within 1-$\sigma$.

\section{Preliminary analysis of the radial-velocity data}

The $K2$ photometry constrained the orbital periods of the transiting planets HIP4178 $d$ and $f$ with 23 different solutions. Since only one transit of planet $e$ has been detected, its orbital period is poorly constrained. In order to determine the orbital period of the outermost planets, an $\ell_{1}$ periodogram\cite{hara2016radial} was computed on the HARPS, HARPS-N, HIRES, and PFS data assuming a jitter of 2\ms\ for the two former instruments and of 4\ms\ for the two latter ones (see Supplementary Table 7) as well as an offset between the instruments. The $\ell_{1}$ periodogram is computed on a grid from 0 to 0.95 cycles per day to avoid the 1 day region, prone to aliasing. 

The results of this periodogram are presented in the Supplementary Figure 4. The main peak is found at a period of 15.5d with a analytical false-alarm probability (FAP)\cite{baluev2008assessing} at the level of $5\times10^{-6}$\ \%. This peak corresponds to the Doppler signal of the transiting planet HIP41378 $b$. The two following strongest peaks are found at periods of 62.2d and at 75.8d. Their analytical FAP are at the level of 0.03\% and 91\%, respectively. These signals support the presence of a non-transiting planet, called $g$, with an orbit of 62.2 days, hence in 2:1 MMR with planet $c$ (the 75.7d signal is the 1-yr alias of the 62.2-d signal). This 62.2-d signal can not be associated to planets $d$, $e$, or $f$ for stability reason$^{\citenum{becker2018discrete}}$. 
 
Although the FAP are greater than 10\%, other periodicities are detected in the $\ell_{1}$ periodogram: 31.5 days corresponding to the transiting planet $c$ as well as a broad peak with two apparent bulges, a small one from $\approx$ 350 to 410 days and a higher one from $\approx$ 410 to 650 days. The latter one corresponds perfectly to the 542-d solution of planet $f$ while the former one is likely related to planet $d$ or $e$. Another peak is detected with a period of 8.2d. Its FAP is however of 10\% and the period does not correspond to the rotation period of the star derived from photometry, nor of a transiting planet and could be due to noise.

A Bayesian General Lomb-Scargle periodogram (BGLS)\cite{mortier2015bgls} was computed from the residuals of the best-fit 6 Keplerian-orbit model (see Supplementary Figure 5) where a peak at 8.2 days is detected. To assess the stability of this periodicity over time, BGLS periodograms were generated for each separate season of data, excluding the first season with the least amount of data. The 8.2d peak is strongly recovered from the season 3 dataset, but absent in season 2 and only marginally present in season 4. Furthermore, by stacking the periodograms\cite{mortier2017stacked}, it is clear that even within season 3, the 8.2d periodicity is unstable over time. This indicates that it is highly unlikely this periodicity is related to a planet.

The $\ell_{1}$ periodogram is only modelling circular orbits without the constraints from the photometry. To further refine the orbital periods of the outermost transiting planets accounting for the transit epochs, the 464 radial-velocity epochs were analysed using a Markov Chain Monte Carlo method from  \texttt{PASTIS}$^{\citenum{diaz2014pastis}}$ with a 5 Keplerian-orbit and a long-term drift model. The ephemerides of planets $b$ and $c$ were fixed to their photometric values. For the planets $d$, $e$, and $f$, a broad prior on their orbital period was used but their epochs of transit were fixed to the one observed during C5. For their orbital eccentricity, a prior was used that corresponds to the eccentricity distribution observed in multiple-planetary systems\cite{van2019orbital}. 

The residuals from the best fit exhibits a periodic signal at $\sim$ 62.2d, corresponding to the low-FAP peak detected by the $\ell_{1}$ periodogram. A 6$^{th}$ planet, $g$, was thus added in the model and the analysis repeated. This non-transiting planet being in 2:1 MMR with planet $c$, it is very likely the source of its large TTVs. The exhaustive list of free parameters are reported in the Supplementary Table 6, together with their prior and posterior distributions. 

Even with a large prior distribution on the orbital period of planet $f$, the MCMC derived a posterior distribution that corresponds perfectly and is compatible only with the 542-d solution from the $K2$ photometry$^{\citenum{becker2018discrete,berardo2019revisiting}}$ (see Supplementary Figure 4). This 542-d solution is also the one that minimised the eccentricity for planet $f$ (Lund et al., submitted). Even if the chains converged towards an orbital period of about 173 days for planet $e$, there is no detection of the planets $d$ and $e$ in the radial velocity data. The data supports no significant drift.

\section{Combined analysis}

The two campaigns (C5 and C18) of the $K2$ mission were analysed together with the 464 epochs from the HARPS, HARPS-N, HIRES, and PFS spectrographs as well as the spectral energy distribution (SED) from the star (see Supplementary Table 1). The analysis was performed with \texttt{PASTIS}$^{\citenum{diaz2014pastis}}$. The photometric data were modelled with \texttt{JKTEBOP} (JKT Eclipsing Binary Orbit Program) software\cite{southworth2008homogeneous}. To account for the long-cadence observations\cite{kipping2010binning} during C5, the photometric model was oversampled at the 1-minute cadence, corresponding to the C18 short-cadence data, before being binned back to the observation sampling to compute the likelihood. The quadratic limb-darkening coefficients were taken from a well adopted theoretical table\cite{claret2011gravity}. The radial velocities were modelled using a 6-planet Keplerian-orbit model. The SED was modelled with the BT-SETTL stellar atmosphere models\cite{allard2012models}. The host star was modelled self-consistently using the Dartmouth stellar-evolution tracks\cite{dotter2008dartmouth}.

The priors on the stellar parameters (\teff, \met, density) are taken from the results of the asteroseismology study of HIP41378 (Lund et al., submitted) and from the $Gaia$ DR2 for the distance to Earth with the parallax corrected for a systematic offset\cite{stassun2018evidence}. 

The priors on the ephemerides of planets $b$ and $c$ as well as the transit epochs for planets $d$, $e$, and $f$ are taken as Gaussian distributions from previous studies$^{\citenum{berardo2019revisiting}}$ with a width enlarged by a factor 100. Since the period of planet $d$ is not detected in the preliminary, radial-velocity analysis, the 278-d solution$^{\citenum{becker2018discrete}}$ that minimises its eccentricity (Lund et al., submitted) was chosen. For planet $e$, a broad prior was used for the orbital period. Since the period of planet $f$ was detected to be 546$\pm$18 days, the 542-d solution$^{\citenum{becker2018discrete}}$ was used as the prior with a Gaussian distribution. Like for planets $b$ and $c$, the width of the Gaussian prior distributions for the orbital periods of planets $d$ and $f$ were enlarged by a factor 100 to limit the bias from the results of previous analysis$^{\citenum{becker2018discrete,berardo2019revisiting}}$. The prior on orbital eccentricity is the one observed for small planets in multiple-system$^{\citenum{van2019orbital}}$. For the other parameters, uninformative priors were used.

The model does not account for gravitational interaction between the planets nor for transit timing variations (TTVs). A photodynamical analysis\cite{almenara2015absolute} would be needed to model the TTVs observed on planet $c$ in the $Spitzer^{\citenum{berardo2019revisiting}}$ and $TESS$ data. This is beyond the scope of this study and will be explored in a following paper.

The exhaustive list of parameters used in the analysis are reported in the Supplementary Table 7 together with their prior distribution. The likelihood was computed assuming that the errors are independent and distributed following:

\begin{equation}
\mathcal{L} = p\left(\mathcal{D} | \theta, \mathcal{I}\right) = \frac{1}{\sqrt{2\pi\left(\sigma_i^{2}+\sigma_j^{2}\right)}}\exp\left[-\frac{1}{2}\frac{\left(x - x_{t}\left(\theta\right)\right)^{2}}{\sigma_i^{2}+\sigma_j^{2}}\right]\ ,
\end{equation}
where $\mathcal{L}$ is the likelihood, $\mathcal{D}$, the data composed by $x$, $\theta$ the parameters, $\mathcal{I}$ the information, $\sigma_i$ the instrumental noise, $\sigma_j$ a jitter noise, and $x_{t}\left(\theta\right)$ the model. 

A total of 96 Markov chains of $3\times10^{5}$ iterations were run, initially starting from the joint prior distribution. Convergence of the chains was tested using a Kolmogorov-Smirnov test to make sure that the chains not only reached the same maximum of likelihood, but that they reached the same posterior distribution$^{\citenum{diaz2014pastis}}$. The converged chains were then merged after removing burn-in phase. Finally, the median and 68.3\% and 95\% credible intervals were derived for each parameter and reported in the Supplementary Table 7. This analysis is similar to the ones performed in previous papers\cite{santerne2018earth,lopez2019exoplanet} reporting low-mass transiting exoplanets observed by $K2$.

In this combined analysis, the radial-velocity signals of planets $b$, $c$, $g$, and $f$ are significantly detected. They have masses ranging between 4.4 \Mearth\ and  12 \Mearth. Except for $g$ that is not transiting, hence its radius is unknown, their bulk densities range from 0.09 \gcm3\ to 2.17 \gcm3, with densities decreasing with increasing orbital distance.

The radial-velocity signatures of planets $d$ and $e$ are not significantly detected in the data. For planet $d$, assuming its orbital period is 278 days, its 95\% credible-interval upper-limit on its mass is only of 4.6 \Mearth, hence a bulk-density upper limit (at 95\% credible interval) of 0.56 \gcm3. 

Unlike in the preliminary analysis, based only on the radial-velocity data, the Markov chains of the combined analysis converged towards an orbital period of 369$\pm$10 days for planet $e$ while only one transit of this planet has been detected so far. This is the result of the high-quality $K2$ C5 light curve together with the stringent stellar-density constraints from asteroseismology and a narrow prior on orbital eccentricity motivated by the architecture of the system. This solution would make planet $e$ close to the 2:3 MMR with planet $f$. With such an orbital period, close to the Earth revolution around the Sun, covering its orbit with spectroscopic observations is impossible from the ground. As a consequence, the mass determined here of $12\pm5$ \Mearth\ is likely affected by some systematics. In particular, the solution might be biased by the few HIRES measurements taken  at high airmass at the beginning of the fourth session (see Figure 1). 

Given that planets $d$, $e$, and $f$ transited within the 80-d duration of C5 and they are close to MMR, every transit of $f$ should be accompanied by a transit of $d$ (2:1 MMR). For the same reason, every two transits of $f$ are escorted by a transit of $e$ (2:3 MMR). The $K2$ campaign C18 was exactly 3 years after C5, hence two orbital periods of planet $f$. That explains why both planets $d$ and $f$ transited which is unlikely given their long orbital periods. For the same reason, a transit of planet $e$ would have been detected if the campaign C18 had started a few weeks earlier.

The radial-velocity amplitudes of all planets in this system are at the level of the instrumental stability. Thus the derived amplitudes might be affected by some unknown systematics. However systematic effects are more likely to mimic a signal than smoothing out a large signal. Since these relatively faint signals correspond to transit ephemerides (except for planet $g$ which is in 2:1 MMR with planet $c$), it is unlikely that these systematics mimicked the radial-velocity signal of a transiting planet, with the right period and transit epoch. If systematics are the origin of some signal, in particular for planet $f$, it means that the planets would have an even lower mass, hence density.

The results of this analysis concerning the stellar properties are compatible by construction with the asteroseismology results (Lund et al., submitted), although 2-$\sigma$ differences are found. The reason for this is likely the fact this analysis self-consistently models the planetary transits and the host star thanks to the astero-profiling density. It might also be the consequence of difference in the stellar evolution tracks, or bias on the \teff\ caused by the SED. 

As a sanity check, the combined analysis was also performed with the PyORBIT code\cite{malavolta2016gaps}. Radial velocities were modelled either with non-interacting Keplerian orbits or using the dynamical integrator TRADES\cite{borsato2014trades}, in two separate runs of the code, while transits were modelled using the {\tt batman} code\cite{kreidberg2015batman}. We used the same priors as in the \texttt{PASTIS} combined analysis (see Supplementary Table 7). Posterior sampling was performed with the affine invariant Markov chain Monte Carlo sampler {\tt emcee}\cite{foreman2013emcee}, employing 300 {\it walkers} over $2\times10^5$ steps. Convergence of the chains and analysis of the results were performed using the same approach as in previous works that made use of PyORBIT. Results are well within 1-$\sigma$ of the ones obtained with the \texttt{PASTIS} analysis. We computed the RV difference between the non-interacting model and the dynamical simulation, using the orbital parameters in Table 1 and for the temporal span of the observations\cite{malavolta2017kepler}, and while this difference shows a variable behaviour with time, the peak-to-peak amplitude is well within 0.1 m/s, i.e. beyond the detectability level.

\end{methods}

\renewcommand{\refname}{Methods references}

\putbib[HIP41378]

\end{bibunit}

\begin{addendum}
 \item We are grateful to Mikkel Lund for sharing the manuscript on the asteroseismic constraints of HIP41378 before its publication. This has been extremely useful for the analyses presented here and to reach these results. The HARPS team is grateful to the HARPS observers who conducted part of the visitor-mode observations at La Silla Observatory:  David Anderson, Nicola Astudillo, Xavier Bonfils, Romina Iba\~nez Bustos, David Ehrenreich, Melissa Hobson, Kristine Lam, Baptiste Lavie, Ester Linder, Felipe Murgas, Lorenzo Pino, Julia Seidel, \& Aur\'elien Wyttenbach.
 
Based on observations collected at the European Organisation for Astronomical Research in the Southern Hemisphere under ESO programmes 198.C-0169 and  0102.C-0171. Based on observations made with the Italian Telescopio Nazionale Galileo (TNG) operated on the island of La Palma by the Fundaci\'on Galileo Galilei of the INAF (Instituto Nazionale di Astrof\'isica) at the Spanish Observatorio del Roque de los Muchachos of the Instituto de Astrof\'isica de Canarias. Some of the data presented herein were obtained at the W. M. Keck Observatory, which is operated as a scientific partnership among the California Institute of Technology, the University of California and the National Aeronautics and Space Administration. The Observatory was made possible by the generous financial support of the W. M. Keck Foundation. The authors wish to recognize and acknowledge the very significant cultural role and reverence that the summit of Maunakea has always had within the indigenous Hawaiian community.  We are most fortunate to have the opportunity to conduct observations from this mountain. This paper includes data gathered with the 6.5 meter Magellan Telescopes located at Las Campanas Observatory, Chile. Based in part on observations made at Observatoire de Haute Provence (CNRS), France. This paper includes data collected by the K2 mission. Funding for the K2 mission is provided by the NASA Science Mission directorate. Based on observations obtained at the Canada-France-Hawaii Telescope (CFHT) which is operated by the National Research Council of Canada, the Institut National des Sciences de l'Univers of the Centre National de la Recherche Scientifique of France, and the University of Hawaii.

This publication makes use of The Data \& Analysis Center for Exoplanets (DACE), which is a facility based at the University of Geneva (CH) dedicated to extrasolar planets data visualisation, exchange and analysis. DACE is a platform of the Swiss National Centre of Competence in Research (NCCR) PlanetS, federating the Swiss expertise in Exoplanet research. The DACE platform is available at https://dace.unige.ch. This research has made use of the NASA Exoplanet Archive, which is operated by the California Institute of Technology, under contract with the National Aeronautics and Space Administration under the Exoplanet Exploration Program. This research has made use of the VizieR catalogue access tool, CDS, Strasbourg, France. The original description of the VizieR service was published in A\&AS 143, 23. This publication makes use of data products from the Two Micron All Sky Survey, which is a joint project of the University of Massachusetts and the Infrared Processing and Analysis Center/California Institute of Technology, funded by the National Aeronautics and Space Administration and the National Science Foundation. This publication makes use of data products from the Wide-field Infrared Survey Explorer, which is a joint project of the University of California, Los Angeles, and the Jet Propulsion Laboratory/California Institute of Technology, funded by the National Aeronautics and Space Administration. This work has made use of data from the European Space Agency (ESA) mission {\it Gaia} (\url{https://www.cosmos.esa.int/gaia}), processed by the {\it Gaia} Data Processing and Analysis Consortium (DPAC, \url{https://www.cosmos.esa.int/web/gaia/dpac/consortium}). Funding for the DPAC has been provided by national institutions, in particular the institutions participating in the {\it Gaia} Multilateral Agreement.

The French group acknowledges financial support from the French Programme National de Plan\'etologie (PNP, INSU). 
The Porto group is supported by the Funda\c{c}\~ao para a Ci\^encia e a Tecnologia (FCT) / MCTES through national funds (PIDDAC) and by FEDER - Fundo Europeu de Desenvolvimento Regional through COMPETE2020 - Programa Operacional Competitividade e Internacionaliza\c{c}\~ao by these grants: \\
UID/FIS/04434/2019; PTDC/FIS-AST/32113/2017 \& POCI-01-0145-FEDER-032113; \\ 
PTDC/FIS-AST/28953/2017 \& POCI-01-0145-FEDER-028953. VAd, SGS, EDM, SCCB acknowledge the support from through Investigador FCT contract nr. IF/00650/2015/CP1273/CT0001, \\
IF/00028/2014/CP1215/CT0002, IF/00849/2015/CP1273/CT0003, IF/01312/2014/CP1215/CT0004 (respectively). SHo acknowledges support by the fellowships PD/BD/128119/2016 funded by FCT (Portugal). ODSD is supported in the form of work contract (DL 57/2016/CP1364/CT0004) funded by national funds through FCT. 
The Geneva group thanks the Swiss National Science Foundation (SNSF) and the Geneva University for their continuous support to the exoplanet researches. This work has been in particular carried out in the frame of the National Centre for Competence in Research ?PlanetS? supported by SNSF. XDu is grateful to the Branco-Weiss Fellowship--Society in Science for its continuous support. This research has been funded by the Spanish State Research Agency (AEI) Projects No.ESP2017-87676-C5-1-R and No. MDM-2017-0737 Unidad de Excelencia ``Mar\'ia de Maeztu'' - Centro de Astrobiolog\'ia (INTA-CSIC). DJA acknowledges support from the STFC via an Ernest Rutherford Fellowship (ST/R00384X/1). DB acknowledges support from an NSERC PGS-D scholarship. DJAB acknowledges support from the UK Space Agency (UKSA). DLP acknowledges support through a Royal Society Wolfson Merit Award. IJMC acknowledges support from the NSF through grant AST-1824644, and from NASA through Caltech/JPL grant RSA-1610091. MRK acknowledges support from the NSF Graduate Research Fellowship, grant No. DGE 1339067. MDa acknowledges financial support from Progetto Premiale 2015 FRONTIERA funding scheme of the Italian Ministry of Education, University, and Research. CDD acknowledges support from the K2 Guest Observer program (80NSSC18K0304 \& 80NSSC19K0099) and the TESS Guest Investigator Program (80NSSC18K1583).  Parts of this work have been supported by the National Aeronautics and Space Administration under grant No. NNX17AB59G issued through the Exoplanets Research Program.  LMa acknowledges support from PLATO ASI-INAF agreement n.2015-019-R.1-2018. AMo acknowledges support from Senior Kavli Institute Fellowships at the University of Cambridge. HPO acknowledges support from Centre National d'Etudes Spatiales (CNES) grant 131425-PLATO. JKT acknowledges that support for this work was provided by NASA through Hubble Fellowship grant HST-HF2-51399.001 awarded by the Space Telescope Science Institute, which is operated by the Association of Universities for Research in Astronomy, Inc., for NASA, under contract NAS5-26555. AVa's work was performed under contract with the California Institute of Technology / Jet Propulsion Laboratory funded by NASA through the Sagan Fellowship Program executed by the NASA Exoplanet Science Institute. CAW would like to acknowledge support from UK Science Technology and Facility Council grant ST/P000312/1. PJW has been supported by STFC through consolidated grants ST/L000733/1 and ST/P000495/1. 
ACC acknowledges support from the Science and Technology Facilities Council (STFC) consolidated grant number ST/R000824/1 and UKSA grant ST/R003203/1. LMW acknowledges support from the Beatrice Watson Parrent Fellowship. JFD acknowledges funding from the European Research Council (ERC) under the H2020 research \& innovation programme (grant agreement \#740651 NewWorlds) 

 \item[Competing Interests] The authors declare that they have no competing financial interests.
 \item[Authors Contributions] ASa led the Analysis Task Force to which LMa, MRK, XDu, AVa, NCH, AMo, TAL are members and wrote the manuscript. CDD coordinated the observations from the radial-velocity instruments. 
HARPS programs CoIs and associated collaborators: VAd, ADJA, DBar, SCCB, DBay, IBo, ASB, FBo, DJAB, MRD, EDM, ODe, RFD, GH\'e, PFi, SHo, JLB, TAL, CLo, OMo, LDN, HPO, DLP, ASa (PI), NCS, SGS, SUd, AVi, PJW. 
HARPS-N CoIs and associated collaborators: ASB, LAB, ACC, RCo, MDa, CDD, XDu, PFi, AGh, AHa, DWL, MLM (chair), CLo, LMa, AFMF, EMo, AMo, CNa, DFP, EPo, KRi, DS\'e, ASo, SUd (chair), AVa, CAW. 
HIRES CoIs, major observers, and data reduction: CDD, MRK, DBe, IJMC (PI), BJF, LAH, AWH, HIs, EAP, ESi, LMW. 
PFS CoIs, builders, and data reduction: FDa, RPB, JDC, SAS, JKT, SXW, JNW.
ESPaDOnS CoIs and data reduction: IBo, DBar, JFD, MRD, JLB, TAL, CMo, ASa (PI), NCS.
 \item[Data Availability Statement] The data that support the plots within this paper and other findings of this study are available from the corresponding author upon reasonable request.
 \item[Correspondence] Correspondence and requests for materials should be addressed to A.~Santerne~(email: alexandre.santerne@lam.fr).
\end{addendum}

\newpage

\begin{figure}
\label{plotRV}
\begin{center}
\includegraphics[width=\textwidth]{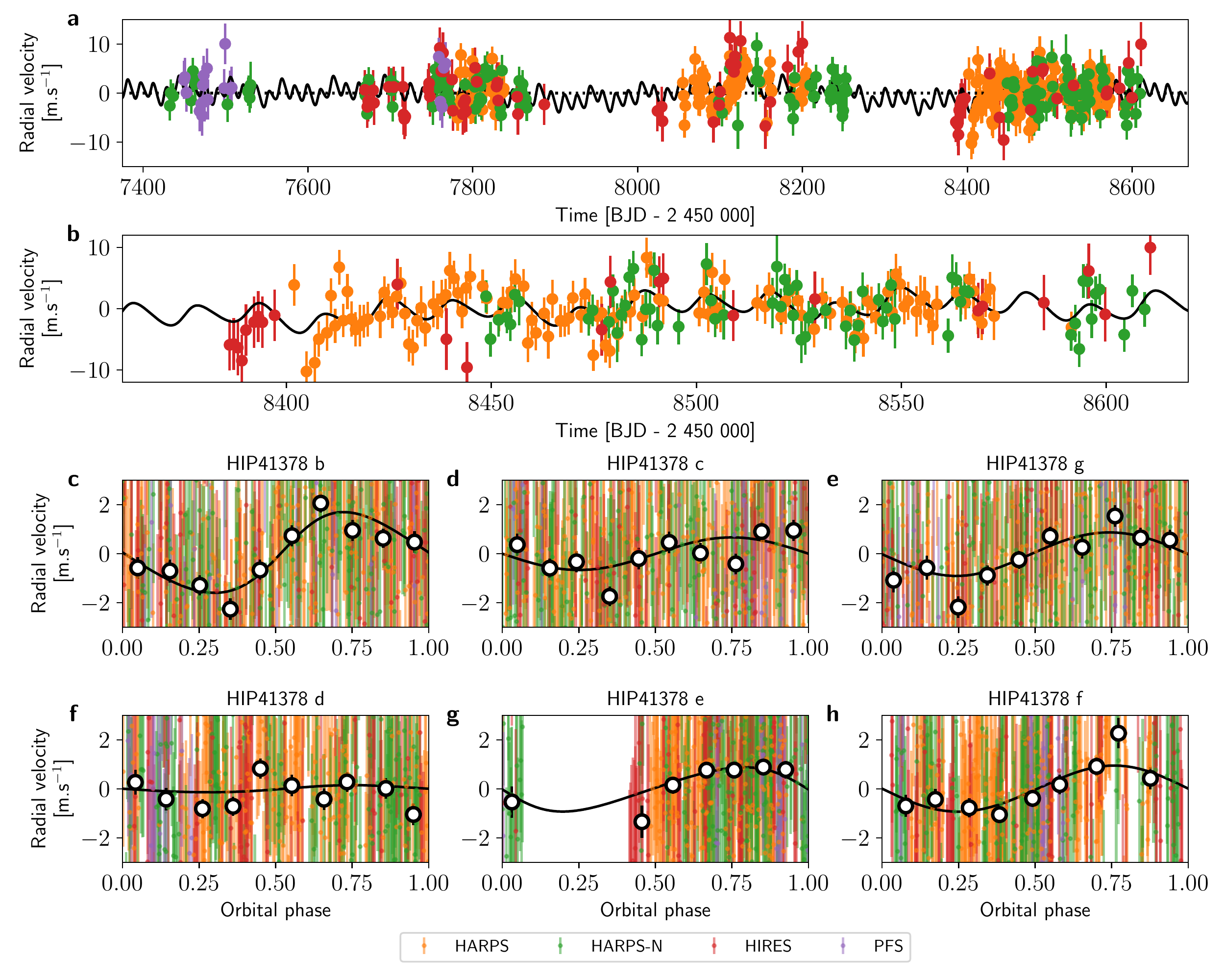}
\caption{\textbf{Radial velocity observations of the star HIP41378.} The different colours correspond to the different instruments: HARPS (orange), HARPS-N (green), HIRES (red), PFS (violet). The black solid line is the best-fit model. Panel \textbf{a} Time series covering the 4 observing seasons. Panel \textbf{b} Zoom of the time series obtained from the last season when HIP41378 was observed every possible night with HARPS. Panels \textbf{c}, \textbf{d}, \textbf{e}, \textbf{f}, \textbf{g}, and \textbf{h} show the radial velocity folded to the phase of the planets $b$, $c$, $g$, $d$, $e$, and $f$ (respectively) once the contribution from the other planets has been removed. Open circles are binned data.}
\end{center}
\end{figure}

\begin{figure}
\begin{center}
\includegraphics[width=\textwidth]{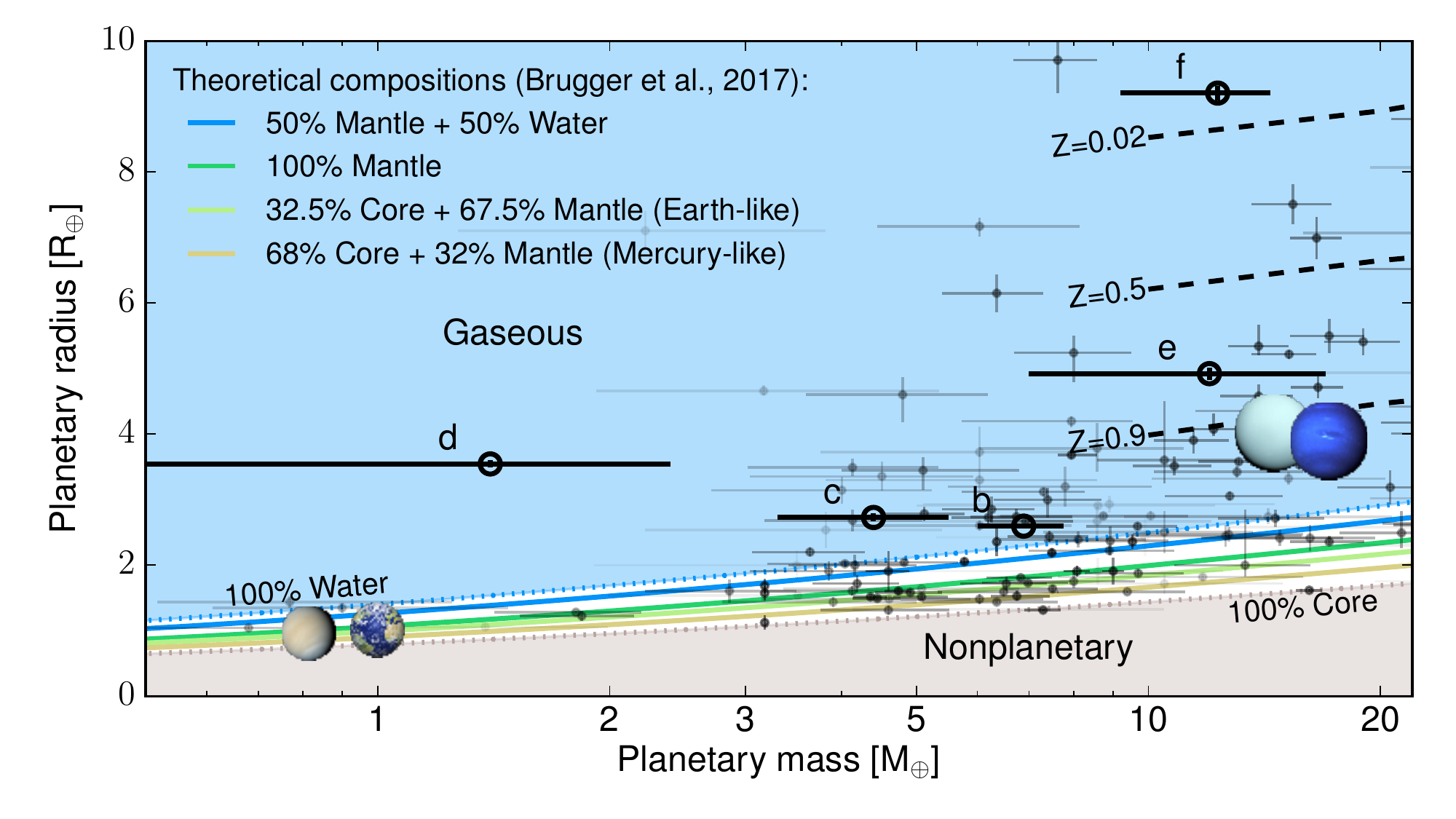}
\label{plotMR}
\caption{\textbf{Mass -- Radius diagram of low-mass exoplanets.} Only planets that have a mass measured with a precision better than 50\% are shown here (source: NASA exoplanet archive). Open circles are the planets in the HIP41378 system. The different colored lines represent possible theoretical compositions for terrestrial worlds$^{\citenum{brugger2017constraints}}$. Objects denser than 100\% metal core are considered as non-planetary objects and those less dense than 100\% water are considered to be gaseous. Black dashed lines are models of giant exoplanets for an age of 3.1 Gyr with metals mass fraction of Z = 0.02 (Solar value), 0.5, and 0.9. The planet on the upper left of HIP41378 $f$ is Kepler-51 $d^{\citenum{masuda2014very}}$.}
\end{center}
\end{figure}

\begin{figure}
\begin{center}
\includegraphics[width=\textwidth]{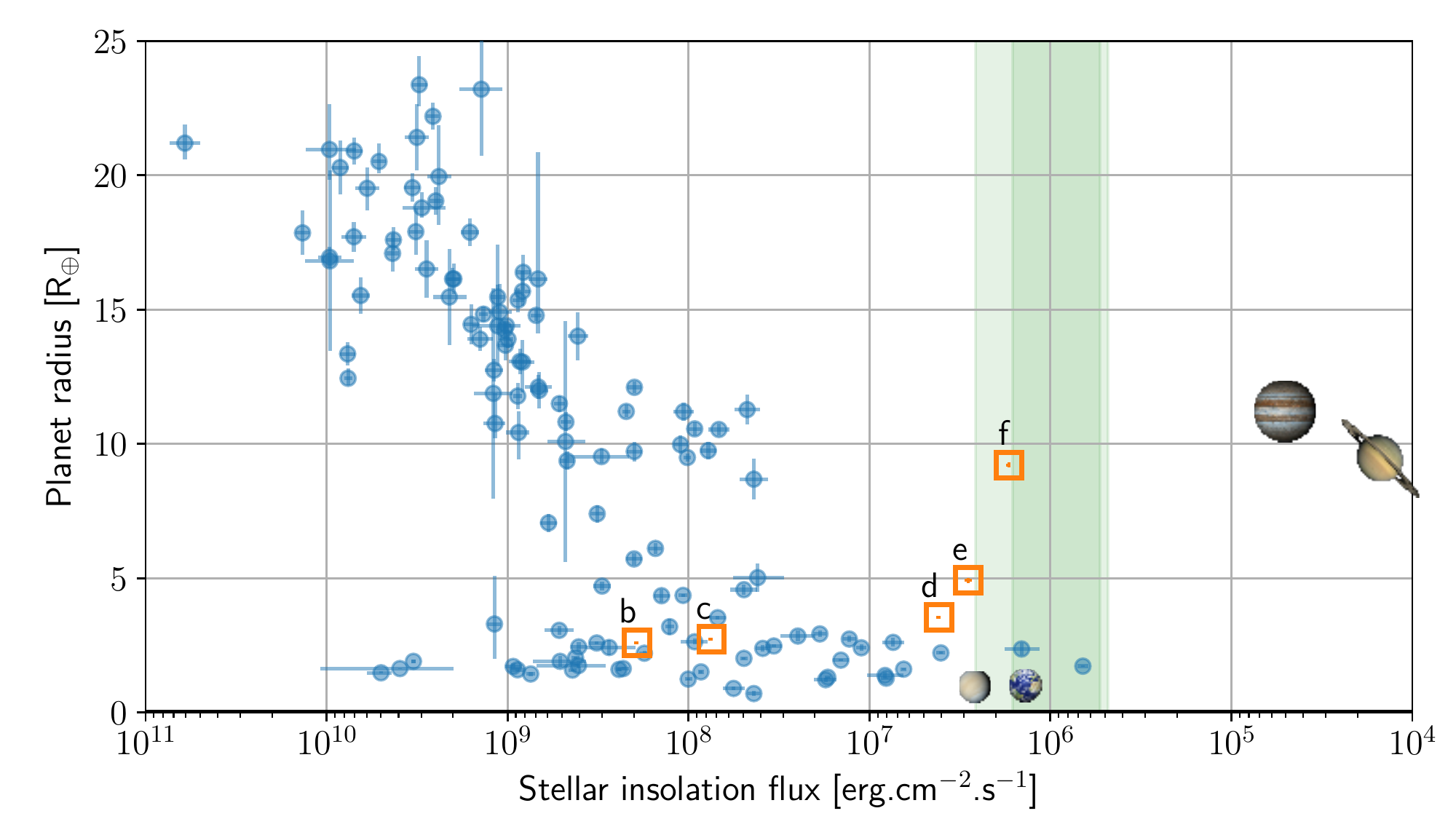}
\label{plotInsolation}
\caption{\textbf{Radius of exoplanets as function of their stellar insolation flux.} Only planets transiting stars that have a magnitude brighter than Ks = 10 are shown here (in blue -- source: NASA exoplanet archive). Venus, the Earth, Jupiter, and Saturn are also displayed for comparison. The five transiting planets in the HIP41378 system are plotted with the orange rectangles. The green zone represent the conservative (green) and optimistic (light green) habitable zone$^{\citenum{kopparapu2013habitable}}$.}
\end{center}
\end{figure}

\newpage

\begin{table}
\centering
\caption{List of the main physical parameters of the HIP41378 planetary system.}
\medskip
\begin{footnotesize}
\begin{tabular}{lccc}
\hline
Parameter & \multicolumn{3}{c}{Median and 68.3\% credible interval}\\
\hline
 & Host star\\
Effective temperature \teff\ [K] & $6320^{_{+60}}_{^{-30}}$\\
Surface gravity \logg\ [cgs] & $4.294\pm0.006$\\
Stellar density $\rho_{\star}$ [$\rho_{\astrosun}$] & $0.563\pm0.006$\\
Iron abundance \met\ [dex] & $-0.10 \pm 0.07$\\
Mass M$_{\star}$\ [\Msun] & $1.16\pm0.04$\\
Radius R$_{\star}$\ [\Rsun] & $1.273\pm0.015$\\
Age $\tau$\ [Gyr] & $3.1\pm0.6$\\
Distance to Earth D [pc] & $103 \pm 2$\\
Rotation velocity \vsini\ [\kms] & $5.6\pm0.5$\\
Rotation period P$_{\rm rot}$ [d] & $6.4\pm0.8$\\
Activity index logR'$_{\rm H\&K}$ & $-4.78\pm0.03$\\
& & & \\
 & Planet $b$ & Planet $c$ & Planet $g$ \\
Period P [d] & $15.57208 \pm 2\times10^{-5}$ & $31.70603\pm 6\times10^{-5}$ & $62.06\pm0.32$\\
Eccentricity e & $0.07\pm0.06$ & $0.04^{_{+0.04}}_{^{-0.03}}$ & $0.06^{_{+0.06}}_{^{-0.04}}$\\
Semi-major axis a [au] & $0.1283 \pm 1.5\times10^{-3}$ & $0.2061 \pm 2.4\times10^{-3}$ & $0.3227\pm.0036$ \\
Inclination i [\degr] & $88.75 \pm 0.13$ & $88.477^{_{+0.035}}_{^{-0.061}}$ & -- \\
Radius R$_{p}$ [\Rearth] & $2.595\pm0.036$ & $2.727\pm0.060$ & -- \\
Mass M$_{p}$ [\Mearth] & $6.89\pm0.88$ & $4.4\pm1.1$ & $7.0\pm1.5^{\ast\ast}$\\
Bulk density $\rho_{p}$ [\gcm3] & $2.17 \pm 0.28$ & $1.19\pm0.30$ &  -- \\
Equilibrium temperature$^{\ast}$ T$_{eq}$ [K] & $959^{_{+9}}_{^{-5}}$ & $757^{_{+7}}_{^{-4}}$ & $605\pm4.7$\\
Stellar insolation flux $S$ [$S_{\oplus}$] & $140^{_{+5}}_{^{-3}}$ & $54^{_{+2}}_{^{-1}}$ & $22.3^{_{+0.8}}_{^{-0.5}}$\\
& & & \\
 & Planet $d$ & Planet $e$ & Planet $f$ \\
Period P [d] &  $278.3618^{\ddag} \pm 5\times10^{-4}$ & $369\pm10$ & $542.07975\pm1.4\times10^{-4}$\\
Eccentricity e & $0.06\pm0.06$ & $0.14\pm0.09$ & $0.004^{_{+0.009}}_{^{-0.003}}$\\
Semi-major axis a [au] & $0.88\pm0.01$ & $1.06^{_{+0.03}}_{^{-0.02}}$ & $1.37 \pm 0.02$\\
Inclination i [\degr] & $89.80 \pm 0.02$ & $89.84^{_{+0.07}}_{^{-0.03}}$ & $89.971^{_{+0.01}}_{^{-0.008}}$\\
Radius R$_{p}$ [\Rearth] & $3.54\pm0.06$ & $4.92\pm0.09$ & $9.2 \pm 0.1$\\
Mass M$_{p}$ [\Mearth] & $<4.6^{\dag}$ & $12\pm5\ (<22^{\dag})$ & $12\pm3$\\
Bulk density $\rho_{p}$ [\gcm3] & $<0.56^{\dag}$ & $0.55\pm0.23\ (<0.82^{\dag})$ &  $0.09\pm0.02$\\
Equilibrium temperature$^{\ast}$ T$_{eq}$ [K] & $367^{_{+3}}_{^{-2}}$ & $335\pm4$ & $294^{_{+3}}_{^{-1}}$\\
Stellar insolation flux $S$ [$S_{\oplus}$] & $3.01^{_{+0.11}}_{^{-0.06}}$ & $2.1\pm0.1$ & $1.24^{_{+0.05}}_{^{-0.02}}$\\
\hline
\multicolumn{4}{l}{$^{\dag}$95\% credible upper limit ; $^{\ddag}$Assumed orbital period ; $^{\ast}$Assuming a zero albedo ; }\\
\multicolumn{4}{l}{$^{\ast\ast}$Assuming an inclination of 88\degr.}\\
\multicolumn{4}{l}{Reference values: \Mearth = 5.9736.10$^{24}$ kg, \Rearth = 6378137 m, \Msun = 1.98842.10$^{30}$ kg, \Rsun = 695508 km,}\\
\multicolumn{4}{l}{1 au = 149597870.7 km, $S_{\oplus}$ = 1366083 erg cm$^{-2}$ s$^{-1}$}
\end{tabular}
\end{footnotesize}
\label{Table1}
\end{table}


\begin{thebibliography}{10}
\expandafter\ifx\csname url\endcsname\relax
  \def\url#1{\texttt{#1}}\fi
\expandafter\ifx\csname urlprefix\endcsname\relax\def\urlprefix{URL }\fi
\providecommand{\bibinfo}[2]{#2}
\providecommand{\eprint}[2][]{\url{#2}}

\bibitem{fortney2013framework}
\bibinfo{author}{Fortney, J.~J.} \emph{et~al.}
\newblock \bibinfo{title}{A framework for characterizing the atmospheres of
  low-mass low-density transiting planets}.
\newblock \emph{\bibinfo{journal}{The Astrophysical Journal}}
  \textbf{\bibinfo{volume}{775}}, \bibinfo{pages}{80} (\bibinfo{year}{2013}).

\bibitem{sing2016continuum}
\bibinfo{author}{Sing, D.~K.} \emph{et~al.}
\newblock \bibinfo{title}{A continuum from clear to cloudy hot-jupiter
  exoplanets without primordial water depletion}.
\newblock \emph{\bibinfo{journal}{Nature}} \textbf{\bibinfo{volume}{529}},
  \bibinfo{pages}{59} (\bibinfo{year}{2016}).

\bibitem{charnay20153d}
\bibinfo{author}{Charnay, B.}, \bibinfo{author}{Meadows, V.},
  \bibinfo{author}{Misra, A.}, \bibinfo{author}{Leconte, J.} \&
  \bibinfo{author}{Arney, G.}
\newblock \bibinfo{title}{3d modeling of gj1214b's atmosphere: formation of
  inhomogeneous high clouds and observational implications}.
\newblock \emph{\bibinfo{journal}{The Astrophysical Journal Letters}}
  \textbf{\bibinfo{volume}{813}}, \bibinfo{pages}{L1} (\bibinfo{year}{2015}).

\bibitem{parmentier2016transitions}
\bibinfo{author}{Parmentier, V.}, \bibinfo{author}{Fortney, J.~J.},
  \bibinfo{author}{Showman, A.~P.}, \bibinfo{author}{Morley, C.} \&
  \bibinfo{author}{Marley, M.~S.}
\newblock \bibinfo{title}{Transitions in the cloud composition of hot
  jupiters}.
\newblock \emph{\bibinfo{journal}{The Astrophysical Journal}}
  \textbf{\bibinfo{volume}{828}}, \bibinfo{pages}{22} (\bibinfo{year}{2016}).

\bibitem{greene2016characterizing}
\bibinfo{author}{Greene, T.~P.} \emph{et~al.}
\newblock \bibinfo{title}{Characterizing transiting exoplanet atmospheres with
  jwst}.
\newblock \emph{\bibinfo{journal}{The Astrophysical Journal}}
  \textbf{\bibinfo{volume}{817}}, \bibinfo{pages}{17} (\bibinfo{year}{2016}).

\bibitem{tinetti2016science}
\bibinfo{author}{Tinetti, G.} \emph{et~al.}
\newblock \bibinfo{title}{The science of ariel (atmospheric remote-sensing
  infrared exoplanet large-survey)}.
\newblock In \emph{\bibinfo{booktitle}{Space Telescopes and Instrumentation
  2016: Optical, Infrared, and Millimeter Wave}}, vol. \bibinfo{volume}{9904},
  \bibinfo{pages}{99041X} (\bibinfo{organization}{International Society for
  Optics and Photonics}, \bibinfo{year}{2016}).

\bibitem{borucki2010kepler}
\bibinfo{author}{Borucki, W.~J.} \emph{et~al.}
\newblock \bibinfo{title}{Kepler planet-detection mission: introduction and
  first results}.
\newblock \emph{\bibinfo{journal}{Science}} \textbf{\bibinfo{volume}{327}},
  \bibinfo{pages}{977--980} (\bibinfo{year}{2010}).

\bibitem{howell2014k2}
\bibinfo{author}{Howell, S.~B.} \emph{et~al.}
\newblock \bibinfo{title}{The k2 mission: characterization and early results}.
\newblock \emph{\bibinfo{journal}{Publications of the Astronomical Society of
  the Pacific}} \textbf{\bibinfo{volume}{126}}, \bibinfo{pages}{398}
  (\bibinfo{year}{2014}).

\bibitem{vanderburg2016five}
\bibinfo{author}{Vanderburg, A.} \emph{et~al.}
\newblock \bibinfo{title}{Five planets transiting a ninth magnitude star}.
\newblock \emph{\bibinfo{journal}{The Astrophysical Journal Letters}}
  \textbf{\bibinfo{volume}{827}}, \bibinfo{pages}{L10} (\bibinfo{year}{2016}).

\bibitem{becker2018discrete}
\bibinfo{author}{Becker, J.~C.} \emph{et~al.}
\newblock \bibinfo{title}{A discrete set of possible transit ephemerides for
  two long-period gas giants orbiting hip 41378}.
\newblock \emph{\bibinfo{journal}{The Astronomical Journal}}
  \textbf{\bibinfo{volume}{157}}, \bibinfo{pages}{19} (\bibinfo{year}{2018}).

\bibitem{berardo2019revisiting}
\bibinfo{author}{Berardo, D.} \emph{et~al.}
\newblock \bibinfo{title}{Revisiting the hip 41378 system with k2 and spitzer}.
\newblock \emph{\bibinfo{journal}{The Astronomical Journal}}
  \textbf{\bibinfo{volume}{157}}, \bibinfo{pages}{185} (\bibinfo{year}{2019}).

\bibitem{ricker2010transiting}
\bibinfo{author}{Ricker, G.~R.} \emph{et~al.}
\newblock \bibinfo{title}{Transiting exoplanet survey satellite (tess)}.
\newblock In \emph{\bibinfo{booktitle}{Bulletin of the American Astronomical
  Society}}, vol.~\bibinfo{volume}{42}, \bibinfo{pages}{459}
  (\bibinfo{year}{2010}).

\bibitem{diaz2014pastis}
\bibinfo{author}{D{\'\i}az, R.~F.} \emph{et~al.}
\newblock \bibinfo{title}{pastis: Bayesian extrasolar planet validation--i.
  general framework, models, and performance}.
\newblock \emph{\bibinfo{journal}{Monthly Notices of the Royal Astronomical
  Society}} \textbf{\bibinfo{volume}{441}}, \bibinfo{pages}{983--1004}
  (\bibinfo{year}{2014}).

\bibitem{brugger2017constraints}
\bibinfo{author}{Brugger, B.}, \bibinfo{author}{Mousis, O.},
  \bibinfo{author}{Deleuil, M.} \& \bibinfo{author}{Deschamps, F.}
\newblock \bibinfo{title}{Constraints on super-earth interiors from stellar
  abundances}.
\newblock \emph{\bibinfo{journal}{The Astrophysical Journal}}
  \textbf{\bibinfo{volume}{850}} (\bibinfo{year}{2017}).

\bibitem{baraffe2008structure}
\bibinfo{author}{Baraffe, I.}, \bibinfo{author}{Chabrier, G.} \&
  \bibinfo{author}{Barman, T.}
\newblock \bibinfo{title}{Structure and evolution of super-earth to
  super-jupiter exoplanets-i. heavy element enrichment in the interior}.
\newblock \emph{\bibinfo{journal}{Astronomy \& Astrophysics}}
  \textbf{\bibinfo{volume}{482}}, \bibinfo{pages}{315--332}
  (\bibinfo{year}{2008}).

\bibitem{mordasini2012characterization}
\bibinfo{author}{Mordasini, C.} \emph{et~al.}
\newblock \bibinfo{title}{Characterization of exoplanets from their
  formation-ii. the planetary mass-radius relationship}.
\newblock \emph{\bibinfo{journal}{Astronomy \& Astrophysics}}
  \textbf{\bibinfo{volume}{547}}, \bibinfo{pages}{A112} (\bibinfo{year}{2012}).

\bibitem{akinsanmi2018detecting}
\bibinfo{author}{Akinsanmi, B.}, \bibinfo{author}{Oshagh, M.},
  \bibinfo{author}{Santos, N.} \& \bibinfo{author}{Barros, S.}
\newblock \bibinfo{title}{Detecting transit signatures of exoplanetary rings
  using soap3. 0}.
\newblock \emph{\bibinfo{journal}{Astronomy \& Astrophysics}}
  \textbf{\bibinfo{volume}{609}}, \bibinfo{pages}{A21} (\bibinfo{year}{2018}).

\bibitem{wang2019dusty}
\bibinfo{author}{Wang, L.} \& \bibinfo{author}{Dai, F.}
\newblock \bibinfo{title}{Dusty outflows in planetary atmospheres:
  Understanding ?super-puffs? and transmission spectra of sub-neptunes}.
\newblock \emph{\bibinfo{journal}{The Astrophysical Journal Letters}}
  \textbf{\bibinfo{volume}{873}}, \bibinfo{pages}{L1} (\bibinfo{year}{2019}).

\bibitem{kopparapu2013habitable}
\bibinfo{author}{Kopparapu, R.~K.} \emph{et~al.}
\newblock \bibinfo{title}{Habitable zones around main-sequence stars: new
  estimates}.
\newblock \emph{\bibinfo{journal}{The Astrophysical Journal}}
  \textbf{\bibinfo{volume}{765}}, \bibinfo{pages}{131} (\bibinfo{year}{2013}).

\bibitem{ronnet2018saturn}
\bibinfo{author}{Ronnet, T.}, \bibinfo{author}{Mousis, O.},
  \bibinfo{author}{Vernazza, P.}, \bibinfo{author}{Lunine, J.~I.} \&
  \bibinfo{author}{Crida, A.}
\newblock \bibinfo{title}{Saturn's formation and early evolution at the origin
  of jupiter's massive moons}.
\newblock \emph{\bibinfo{journal}{The Astronomical Journal}}
  \textbf{\bibinfo{volume}{155}}, \bibinfo{pages}{224} (\bibinfo{year}{2018}).

\bibitem{pepe2010espresso}
\bibinfo{author}{Pepe, F.~A.} \emph{et~al.}
\newblock \bibinfo{title}{Espresso: the echelle spectrograph for rocky
  exoplanets and stable spectroscopic observations}.
\newblock In \emph{\bibinfo{booktitle}{Ground-based and Airborne
  Instrumentation for Astronomy III}}, vol. \bibinfo{volume}{7735},
  \bibinfo{pages}{77350F} (\bibinfo{organization}{International Society for
  Optics and Photonics}, \bibinfo{year}{2010}).

\bibitem{masuda2014very}
\bibinfo{author}{Masuda, K.}
\newblock \bibinfo{title}{Very low density planets around kepler-51 revealed
  with transit timing variations and an anomaly similar to a planet-planet
  eclipse event}.
\newblock \emph{\bibinfo{journal}{The Astrophysical Journal}}
  \textbf{\bibinfo{volume}{783}}, \bibinfo{pages}{53} (\bibinfo{year}{2014}).

\bibitem{jontof2014kepler}
\bibinfo{author}{Jontof-Hutter, D.}, \bibinfo{author}{Lissauer, J.~J.},
  \bibinfo{author}{Rowe, J.~F.} \& \bibinfo{author}{Fabrycky, D.~C.}
\newblock \bibinfo{title}{Kepler-79's low density planets}.
\newblock \emph{\bibinfo{journal}{The Astrophysical Journal}}
  \textbf{\bibinfo{volume}{785}}, \bibinfo{pages}{15} (\bibinfo{year}{2014}).

\bibitem{libby2019featureless}
\bibinfo{author}{Libby-Roberts, J.~E.} \emph{et~al.}
\newblock \bibinfo{title}{The featureless transmission spectra of two
  super-puff planets}.
\newblock \emph{\bibinfo{journal}{arXiv preprint arXiv:1910.12988}}
  (\bibinfo{year}{2019}).

\bibitem{rauer2014plato}
\bibinfo{author}{Rauer, H.} \emph{et~al.}
\newblock \bibinfo{title}{The plato 2.0 mission}.
\newblock \emph{\bibinfo{journal}{Experimental Astronomy}}
  \textbf{\bibinfo{volume}{38}}, \bibinfo{pages}{249--330}
  (\bibinfo{year}{2014}).

\end{thebibliography}


\begin{thebibliography}{10}
\expandafter\ifx\csname url\endcsname\relax
  \def\url#1{\texttt{#1}}\fi
\expandafter\ifx\csname urlprefix\endcsname\relax\def\urlprefix{URL }\fi
\providecommand{\bibinfo}[2]{#2}
\providecommand{\eprint}[2][]{\url{#2}}

\bibitem{vanderburg2014technique}
\bibinfo{author}{Vanderburg, A.} \& \bibinfo{author}{Johnson, J.~A.}
\newblock \bibinfo{title}{A technique for extracting highly precise photometry
  for the two-wheeled kepler mission}.
\newblock \emph{\bibinfo{journal}{Publications of the Astronomical Society of
  the Pacific}} \textbf{\bibinfo{volume}{126}}, \bibinfo{pages}{948}
  (\bibinfo{year}{2014}).

\bibitem{vanderburg2016planetary}
\bibinfo{author}{Vanderburg, A.} \emph{et~al.}
\newblock \bibinfo{title}{Planetary candidates from the first year of the k2
  mission}.
\newblock \emph{\bibinfo{journal}{The Astrophysical Journal Supplement Series}}
  \textbf{\bibinfo{volume}{222}}, \bibinfo{pages}{14} (\bibinfo{year}{2016}).

\bibitem{phase2003setting}
\bibinfo{author}{{Mayor}, M.} \emph{et~al.}
\newblock \bibinfo{title}{{Setting New Standards with {HARPS}}}.
\newblock \emph{\bibinfo{journal}{The Messenger}}
  \textbf{\bibinfo{volume}{114}}, \bibinfo{pages}{20--24}
  (\bibinfo{year}{2003}).

\bibitem{cosentino2012harps}
\bibinfo{author}{Cosentino, R.} \emph{et~al.}
\newblock \bibinfo{title}{Harps-n: the new planet hunter at tng}.
\newblock In \emph{\bibinfo{booktitle}{Ground-based and Airborne
  Instrumentation for Astronomy IV}}, vol. \bibinfo{volume}{8446},
  \bibinfo{pages}{84461V} (\bibinfo{organization}{International Society for
  Optics and Photonics}, \bibinfo{year}{2012}).

\bibitem{baranne1996elodie}
\bibinfo{author}{Baranne, A.} \emph{et~al.}
\newblock \bibinfo{title}{Elodie: A spectrograph for accurate radial velocity
  measurements}.
\newblock \emph{\bibinfo{journal}{Astronomy and Astrophysics Supplement
  Series}} \textbf{\bibinfo{volume}{119}}, \bibinfo{pages}{373--390}
  (\bibinfo{year}{1996}).

\bibitem{pepe2002coralie}
\bibinfo{author}{Pepe, F.} \emph{et~al.}
\newblock \bibinfo{title}{The coralie survey for southern extra-solar planets
  vii-two short-period saturnian companions to hd 108147 and hd 168746}.
\newblock \emph{\bibinfo{journal}{Astronomy \& Astrophysics}}
  \textbf{\bibinfo{volume}{388}}, \bibinfo{pages}{632--638}
  (\bibinfo{year}{2002}).

\bibitem{dumusque2011planetary}
\bibinfo{author}{Dumusque, X.}, \bibinfo{author}{Udry, S.},
  \bibinfo{author}{Lovis, C.}, \bibinfo{author}{Santos, N.~C.} \&
  \bibinfo{author}{Monteiro, M.}
\newblock \bibinfo{title}{Planetary detection limits taking into account
  stellar noise-i. observational strategies to reduce stellar oscillation and
  granulation effects}.
\newblock \emph{\bibinfo{journal}{Astronomy \& Astrophysics}}
  \textbf{\bibinfo{volume}{525}}, \bibinfo{pages}{A140} (\bibinfo{year}{2011}).

\bibitem{vogt1994hires}
\bibinfo{author}{Vogt, S.~S.} \emph{et~al.}
\newblock \bibinfo{title}{Hires: the high-resolution echelle spectrometer on
  the keck 10-m telescope}.
\newblock In \emph{\bibinfo{booktitle}{Instrumentation in Astronomy VIII}},
  vol. \bibinfo{volume}{2198}, \bibinfo{pages}{362--375}
  (\bibinfo{organization}{International Society for Optics and Photonics},
  \bibinfo{year}{1994}).

\bibitem{butler1996attaining}
\bibinfo{author}{Butler, R.~P.} \emph{et~al.}
\newblock \bibinfo{title}{Attaining doppler precision of 3 m s-1}.
\newblock \emph{\bibinfo{journal}{Publications of the Astronomical Society of
  the Pacific}} \textbf{\bibinfo{volume}{108}}, \bibinfo{pages}{500}
  (\bibinfo{year}{1996}).

\bibitem{howard2010occurrence}
\bibinfo{author}{Howard, A.~W.} \emph{et~al.}
\newblock \bibinfo{title}{The occurrence and mass distribution of close-in
  super-earths, neptunes, and jupiters}.
\newblock \emph{\bibinfo{journal}{Science}} \textbf{\bibinfo{volume}{330}},
  \bibinfo{pages}{653--655} (\bibinfo{year}{2010}).

\bibitem{crane2006carnegie}
\bibinfo{author}{Crane, J.~D.}, \bibinfo{author}{Shectman, S.~A.} \&
  \bibinfo{author}{Butler, R.~P.}
\newblock \bibinfo{title}{The carnegie planet finder spectrograph}.
\newblock In \emph{\bibinfo{booktitle}{Ground-based and Airborne
  Instrumentation for Astronomy}}, vol. \bibinfo{volume}{6269},
  \bibinfo{pages}{626931} (\bibinfo{organization}{International Society for
  Optics and Photonics}, \bibinfo{year}{2006}).

\bibitem{donati2006espadons}
\bibinfo{author}{Donati, J.-F.}, \bibinfo{author}{Catala, C.},
  \bibinfo{author}{Landstreet, J.} \& \bibinfo{author}{Petit, P.}
\newblock \bibinfo{title}{Espadons: the new generation stellar
  spectro-polarimeter. performances and first results}.
\newblock In \emph{\bibinfo{booktitle}{Solar Polarization 4}}, vol.
  \bibinfo{volume}{358}, \bibinfo{pages}{362} (\bibinfo{year}{2006}).

\bibitem{donati2008magnetic}
\bibinfo{author}{Donati, J.-F.} \emph{et~al.}
\newblock \bibinfo{title}{Magnetic cycles of the planet-hosting star $\tau$
  bootis}.
\newblock \emph{\bibinfo{journal}{Monthly Notices of the Royal Astronomical
  Society}} \textbf{\bibinfo{volume}{385}}, \bibinfo{pages}{1179--1185}
  (\bibinfo{year}{2008}).

\bibitem{fares2009magnetic}
\bibinfo{author}{Fares, R.} \emph{et~al.}
\newblock \bibinfo{title}{Magnetic cycles of the planet-hosting star $\tau$
  bootis--ii. a second magnetic polarity reversal}.
\newblock \emph{\bibinfo{journal}{Monthly Notices of the Royal Astronomical
  Society}} \textbf{\bibinfo{volume}{398}}, \bibinfo{pages}{1383--1391}
  (\bibinfo{year}{2009}).

\bibitem{mengel2016evolving}
\bibinfo{author}{Mengel, M.} \emph{et~al.}
\newblock \bibinfo{title}{The evolving magnetic topology of $\tau$ bo{\"o}tis}.
\newblock \emph{\bibinfo{journal}{Monthly Notices of the Royal Astronomical
  Society}} \textbf{\bibinfo{volume}{459}}, \bibinfo{pages}{4325--4342}
  (\bibinfo{year}{2016}).

\bibitem{noyes1984relation}
\bibinfo{author}{Noyes, R.}, \bibinfo{author}{Weiss, N.} \&
  \bibinfo{author}{Vaughan, A.}
\newblock \bibinfo{title}{The relation between stellar rotation rate and
  activity cycle periods}.
\newblock \emph{\bibinfo{journal}{The Astrophysical Journal}}
  \textbf{\bibinfo{volume}{287}}, \bibinfo{pages}{769--773}
  (\bibinfo{year}{1984}).

\bibitem{lovis2011harps}
\bibinfo{author}{Lovis, C.} \emph{et~al.}
\newblock \bibinfo{title}{The harps search for southern extra-solar planets.
  xxxi. magnetic activity cycles in solar-type stars: statistics and impact on
  precise radial velocities}.
\newblock \emph{\bibinfo{journal}{arXiv preprint arXiv:1107.5325}}
  (\bibinfo{year}{2011}).

\bibitem{donahue1993surface}
\bibinfo{author}{Donahue, R.~A.}
\newblock \bibinfo{title}{Surface differential rotation in a sample of cool
  dwarf stars}.
\newblock \emph{\bibinfo{journal}{Publications of the Astronomical Society of
  the Pacific}} \textbf{\bibinfo{volume}{105}}, \bibinfo{pages}{804}
  (\bibinfo{year}{1993}).

\bibitem{wright2004chromospheric}
\bibinfo{author}{Wright, J.~T.}, \bibinfo{author}{Marcy, G.~W.},
  \bibinfo{author}{Butler, R.~P.} \& \bibinfo{author}{Vogt, S.~S.}
\newblock \bibinfo{title}{Chromospheric ca ii emission in nearby f, g, k, and m
  stars}.
\newblock \emph{\bibinfo{journal}{The Astrophysical Journal Supplement Series}}
  \textbf{\bibinfo{volume}{152}}, \bibinfo{pages}{261} (\bibinfo{year}{2004}).

\bibitem{mamajek2008improved}
\bibinfo{author}{Mamajek, E.~E.} \& \bibinfo{author}{Hillenbrand, L.~A.}
\newblock \bibinfo{title}{Improved age estimation for solar-type dwarfs using
  activity-rotation diagnostics}.
\newblock \emph{\bibinfo{journal}{The Astrophysical Journal}}
  \textbf{\bibinfo{volume}{687}}, \bibinfo{pages}{1264} (\bibinfo{year}{2008}).

\bibitem{hara2016radial}
\bibinfo{author}{Hara, N.~C.}, \bibinfo{author}{Bou{\'e}, G.},
  \bibinfo{author}{Laskar, J.} \& \bibinfo{author}{Correia, A.}
\newblock \bibinfo{title}{Radial velocity data analysis with compressed sensing
  techniques}.
\newblock \emph{\bibinfo{journal}{Monthly Notices of the Royal Astronomical
  Society}} \textbf{\bibinfo{volume}{464}}, \bibinfo{pages}{1220--1246}
  (\bibinfo{year}{2016}).

\bibitem{baluev2008assessing}
\bibinfo{author}{Baluev, R.~V.}
\newblock \bibinfo{title}{Assessing the statistical significance of periodogram
  peaks}.
\newblock \emph{\bibinfo{journal}{Monthly Notices of the Royal Astronomical
  Society}} \textbf{\bibinfo{volume}{385}}, \bibinfo{pages}{1279--1285}
  (\bibinfo{year}{2008}).

\bibitem{mortier2015bgls}
\bibinfo{author}{Mortier, A.}, \bibinfo{author}{Faria, J.},
  \bibinfo{author}{Correia, C.}, \bibinfo{author}{Santerne, A.} \&
  \bibinfo{author}{Santos, N.}
\newblock \bibinfo{title}{Bgls: A bayesian formalism for the generalised
  lomb-scargle periodogram}.
\newblock \emph{\bibinfo{journal}{Astronomy \& Astrophysics}}
  \textbf{\bibinfo{volume}{573}}, \bibinfo{pages}{A101} (\bibinfo{year}{2015}).

\bibitem{mortier2017stacked}
\bibinfo{author}{Mortier, A.} \& \bibinfo{author}{Cameron, A.~C.}
\newblock \bibinfo{title}{Stacked bayesian general lomb-scargle periodogram:
  Identifying stellar activity signals}.
\newblock \emph{\bibinfo{journal}{Astronomy \& Astrophysics}}
  \textbf{\bibinfo{volume}{601}}, \bibinfo{pages}{A110} (\bibinfo{year}{2017}).

\bibitem{van2019orbital}
\bibinfo{author}{Van~Eylen, V.} \emph{et~al.}
\newblock \bibinfo{title}{The orbital eccentricity of small planet systems}.
\newblock \emph{\bibinfo{journal}{The Astronomical Journal}}
  \textbf{\bibinfo{volume}{157}}, \bibinfo{pages}{61} (\bibinfo{year}{2019}).

\bibitem{southworth2008homogeneous}
\bibinfo{author}{Southworth, J.}
\newblock \bibinfo{title}{Homogeneous studies of transiting extrasolar
  planets--i. light-curve analyses}.
\newblock \emph{\bibinfo{journal}{Monthly Notices of the Royal Astronomical
  Society}} \textbf{\bibinfo{volume}{386}}, \bibinfo{pages}{1644--1666}
  (\bibinfo{year}{2008}).

\bibitem{kipping2010binning}
\bibinfo{author}{Kipping, D.~M.}
\newblock \bibinfo{title}{Binning is sinning: morphological light-curve
  distortions due to finite integration time}.
\newblock \emph{\bibinfo{journal}{Monthly Notices of the Royal Astronomical
  Society}} \textbf{\bibinfo{volume}{408}}, \bibinfo{pages}{1758--1769}
  (\bibinfo{year}{2010}).

\bibitem{claret2011gravity}
\bibinfo{author}{Claret, A.} \& \bibinfo{author}{Bloemen, S.}
\newblock \bibinfo{title}{Gravity and limb-darkening coefficients for the
  kepler, corot, spitzer, uvby, ubvrijhk, and sloan photometric systems}.
\newblock \emph{\bibinfo{journal}{Astronomy \& Astrophysics}}
  \textbf{\bibinfo{volume}{529}}, \bibinfo{pages}{A75} (\bibinfo{year}{2011}).

\bibitem{allard2012models}
\bibinfo{author}{Allard, F.}, \bibinfo{author}{Homeier, D.} \&
  \bibinfo{author}{Freytag, B.}
\newblock \bibinfo{title}{Models of very-low-mass stars, brown dwarfs and
  exoplanets}.
\newblock \emph{\bibinfo{journal}{Philosophical Transactions of the Royal
  Society A: Mathematical, Physical and Engineering Sciences}}
  \textbf{\bibinfo{volume}{370}}, \bibinfo{pages}{2765--2777}
  (\bibinfo{year}{2012}).

\bibitem{dotter2008dartmouth}
\bibinfo{author}{Dotter, A.} \emph{et~al.}
\newblock \bibinfo{title}{The dartmouth stellar evolution database}.
\newblock \emph{\bibinfo{journal}{The Astrophysical Journal Supplement Series}}
  \textbf{\bibinfo{volume}{178}}, \bibinfo{pages}{89} (\bibinfo{year}{2008}).

\bibitem{stassun2018evidence}
\bibinfo{author}{Stassun, K.~G.} \& \bibinfo{author}{Torres, G.}
\newblock \bibinfo{title}{Evidence for a systematic offset of- 80 $\mu$as in
  the gaia dr2 parallaxes}.
\newblock \emph{\bibinfo{journal}{The Astrophysical Journal}}
  \textbf{\bibinfo{volume}{862}}, \bibinfo{pages}{61} (\bibinfo{year}{2018}).

\bibitem{almenara2015absolute}
\bibinfo{author}{Almenara, J.} \emph{et~al.}
\newblock \bibinfo{title}{Absolute masses and radii determination in
  multiplanetary systems without stellar models}.
\newblock \emph{\bibinfo{journal}{Monthly Notices of the Royal Astronomical
  Society}} \textbf{\bibinfo{volume}{453}}, \bibinfo{pages}{2644--2652}
  (\bibinfo{year}{2015}).

\bibitem{santerne2018earth}
\bibinfo{author}{Santerne, A.} \emph{et~al.}
\newblock \bibinfo{title}{An earth-sized exoplanet with a mercury-like
  composition}.
\newblock \emph{\bibinfo{journal}{Nature Astronomy}}
  \textbf{\bibinfo{volume}{2}}, \bibinfo{pages}{393} (\bibinfo{year}{2018}).

\bibitem{lopez2019exoplanet}
\bibinfo{author}{Lopez, T.} \emph{et~al.}
\newblock \bibinfo{title}{Exoplanet characterisation in the longest known
  resonant chain: the k2-138 system seen by harps}.
\newblock \emph{\bibinfo{journal}{Astronomy \& Astrophysics}}
  \textbf{\bibinfo{volume}{631}}, \bibinfo{pages}{A90} (\bibinfo{year}{2019}).

\bibitem{malavolta2016gaps}
\bibinfo{author}{Malavolta, L.} \emph{et~al.}
\newblock \bibinfo{title}{The gaps programme with harps-n at tng-xi. pr 0211 in
  m 44: the first multi-planet system in an open cluster}.
\newblock \emph{\bibinfo{journal}{Astronomy \& Astrophysics}}
  \textbf{\bibinfo{volume}{588}}, \bibinfo{pages}{A118} (\bibinfo{year}{2016}).

\bibitem{borsato2014trades}
\bibinfo{author}{Borsato, L.} \emph{et~al.}
\newblock \bibinfo{title}{Trades: A new software to derive orbital parameters
  from observed transit times and radial velocities-revisiting kepler-11 and
  kepler-9}.
\newblock \emph{\bibinfo{journal}{Astronomy \& Astrophysics}}
  \textbf{\bibinfo{volume}{571}}, \bibinfo{pages}{A38} (\bibinfo{year}{2014}).

\bibitem{kreidberg2015batman}
\bibinfo{author}{Kreidberg, L.}
\newblock \bibinfo{title}{batman: Basic transit model calculation in python}.
\newblock \emph{\bibinfo{journal}{Publications of the Astronomical Society of
  the Pacific}} \textbf{\bibinfo{volume}{127}}, \bibinfo{pages}{1161}
  (\bibinfo{year}{2015}).

\bibitem{foreman2013emcee}
\bibinfo{author}{Foreman-Mackey, D.}, \bibinfo{author}{Hogg, D.~W.},
  \bibinfo{author}{Lang, D.} \& \bibinfo{author}{Goodman, J.}
\newblock \bibinfo{title}{emcee: the mcmc hammer}.
\newblock \emph{\bibinfo{journal}{Publications of the Astronomical Society of
  the Pacific}} \textbf{\bibinfo{volume}{125}}, \bibinfo{pages}{306}
  (\bibinfo{year}{2013}).

\bibitem{malavolta2017kepler}
\bibinfo{author}{Malavolta, L.} \emph{et~al.}
\newblock \bibinfo{title}{The kepler-19 system: a thick-envelope super-earth
  with two neptune-mass companions characterized using radial velocities and
  transit timing variations}.
\newblock \emph{\bibinfo{journal}{The Astronomical Journal}}
  \textbf{\bibinfo{volume}{153}}, \bibinfo{pages}{224} (\bibinfo{year}{2017}).

\end{thebibliography}
\end{document}